\PassOptionsToPackage{table,xcdraw}{xcolor}
\documentclass[sigconf,nonacm,10pt]{acmart}

\usepackage{caption}
\usepackage{subcaption}
\usepackage{bbm,amsmath}
\usepackage[ruled,linesnumbered]{algorithm2e}
\usepackage{multirow}
\usepackage{makecell}
\usepackage{tabularx}
\usepackage{circledtext}
\usepackage{array}

\newcolumntype{Y}[1]{%
  >{\hsize=#1\hsize\linewidth=\hsize
    \centering\arraybackslash}X%
}

\renewcommand\footnotetextcopyrightpermission[1]{}
\AtBeginDocument{%
  }

\begin{document}

\title[Zero-Knowledge Remote Adversarial Attack against Wi-Fi-based HAR for Privacy Protection]{Zero-Knowledge Remote Adversarial Attack against Wi-Fi-based Human Activity Recognition for Privacy Protection\vspace{-0.3cm}}

\author{\large Byungjun Kim$^{\star}$, Amogh Panchagatti$^{\diamond}$, Peter Gerstoft$^{\diamond}$, Xinyu Zhang$^{\diamond}$, Minsung Kim$^{\star}$}

\affiliation{%
 \institution{\small $^{\star}$Rutgers University, $^{\diamond}$UCSD}
 \country{}
 }

\newcommand{\systemname}{{\sf{GRAW}}}

\newcommand{\HH}{\mathbf{H}}
\renewcommand{\AA}{\mathbf{A}}
\newcommand{\BB}{\mathbf{B}}
\newcommand{\HHH}{\mathcal{H}}
\newcommand{\AAA}{\mathcal{A}}
\newcommand{\BBB}{\mathcal{B}}

\newcommand{\nRX}{N_{\rm{RX}}}
\newcommand{\nTX}{N_{\rm{TX}}}
\newcommand{\nSC}{N_{\rm{SC}}}

\newcommand{\fC}{f_\textrm{C}}
\newcommand{\fCsur}{f'_\textrm{C}}
\newcommand{\fLSTM}{f_\textrm{LSTM}}
\newcommand{\fLSTMsur}{f'_\textrm{LSTM}}
\newcommand{\fLSTMsurAug}{f''_\textrm{LSTM}}

\newcommand{\TAR}{{\sf{TAR}}}
\newcommand{\JAR}{{\sf{JAR}}}
\newcommand{\RUAR}{{\sf{RUAR}}}

\newcommand{\WiCAM}{{\sf{WiCAM}}}
\newcommand{\CW}{{\sf{C\&W}}}
\newcommand{\ISWARS}{{\sf{IS-WARS}}}
\newcommand{\AAEN}{{\sf{AAEN}}}

\newcommand{\parahead}[1]{\vspace{2pt plus 0pt minus 2pt}\noindent{\bfseries #1}}
\newcommand{\parabreak}{\vspace*{1.00ex minus 0.25ex}\noindent}

\definecolor{LightGreen}{rgb}{0.88,1,0.88}
\definecolor{DarkGreen}{rgb}{0.0,0.4,0.13}
\definecolor{LightOrange}{rgb}{1,0.85,0.8}
\definecolor{LightYellow}{rgb}{1,1.00,0.5}
\definecolor{LightRed}{rgb}{1,0.80,0.80}

\renewcommand{\shortauthors}{Byungjun Kim, Amogh Panchagatti, Peter Gerstoft, Xinyu Zhang, Minsung Kim}

\begin{abstract}
The growing capability of Wi-Fi devices to identify human activities using channel state information (CSI) raises privacy concerns. To counter this threat, we propose \systemname{}, an adversary system—acting as a privacy defender—that degrades the human activity recognition (HAR) system at the user device by perturbing the router's signals that the device uses to estimate CSI.
\systemname{} employs \emph{generative adversarial imitation learning} (GAIL) to construct perturbation signals, and thereby eliminates the need for any information on the target HAR systems and their inputs (\emph{i.e.,} zero-knowledge operation). We evaluate \systemname{} against seven representative HAR models, using datasets collected in five environments, including our own dataset. We observe that \systemname{} is the only remote attack scheme that degrades every tested HAR model to a random-selection level. At the same perturbation level, \systemname{} achieves an attack success ratio up to 76.7\% higher than comparison methods, while maintaining over 99\% packet success rate on regular Wi-Fi communication. 
We demonstrate the feasibility of \systemname{} through real-time, over-the-air experiments with software-defined radios.  
\end{abstract}





\maketitle

\begin{figure}
\centering
\includegraphics[width=.83\columnwidth]{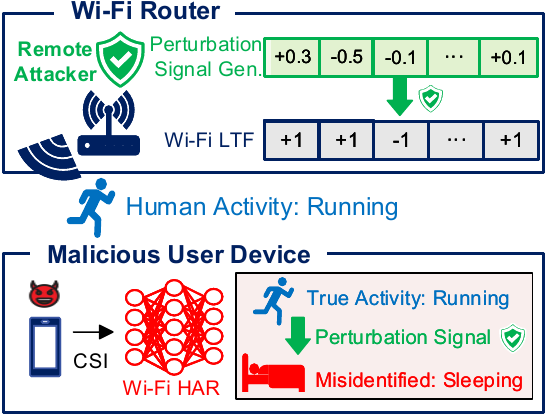}
\vspace{-0.2cm}
\caption{Remote adversarial attack against Wi-Fi HAR. The adversary in the Wi-Fi router attacks neural network-based HAR in the malicious device by adding perturbation signals to Wi-Fi LTF preambles from the router side, aiming to disable or degrade HAR.}\label{fig:sysModel}\vspace{-0.4cm}
\end{figure}

\section{Introduction}
\label{sec:intro}

As Wi-Fi devices become prevalent indoors, Wi-Fi sensing using \emph{channel state information} (CSI) has been used for various purposes, including indoor localization~\cite{wang2018deep, tong2021wi}, radio fingerprinting~\cite{meneghello2022deepcsi, chapre2014csi}, and human mesh construction with millimeter-wave Wi-Fi~\cite{wang2022wi}. One emerging application of Wi-Fi sensing is \emph{human activity recognition} (HAR)~\cite{muaaz2020wiwehar, xi2015device}, which aims to identify human activities using Wi-Fi CSI sequence and neural networks. The enhanced capabilities of modern Wi-Fi HAR systems, while beneficial for legitimate applications, also pose serious privacy threats ~\cite{zhang2020understanding, liu2023time} due to unauthorized surveillance capabilities, such as unintended motion sensing~\cite{zhu2018tu} and keystroke eavesdropping~\cite{hu2023password}.

Such unauthorized HAR may be performed on malicious \emph{HAR devices}---user devices that recognize activities from CSI without consent. To protect user privacy from these devices, \emph{adversarial attacks} have been extensively studied, where the attacker deliberately degrades the HAR performance of their neural networks.
Previous research has explored the direct manipulation of inputs on HAR devices, such as HAR classifier’s loss functions and CSI sequences~\cite{xu2022wicam,zhang2020understanding}; this input manipulation approach is called a \emph{digital attack}. In digital attacks, simple methods like the fast gradient method (FGM) have proven effective~\cite{szegedy2013intriguing}.
However, digital-attack scenarios assume adversarial attacker systems have direct access to HAR inputs at the Wi-Fi HAR device, which is impractical; for example, the used CSI sequence is estimated within the HAR device, beyond the attacker's control.

To address this limitation, recent work~\cite{li2024practical,huang2021wars} has investigated the manipulation of Wi-Fi preamble signals 
to remotely disrupt HAR systems on the receiving end of the communication link, which we refer to as a \emph{remote attack}.
In Wi-Fi systems, a router sends a \emph{Long Training Field} (LTF) signal, \emph{i.e.,} a preamble known to both the router and users, to a user device, and the device estimates the CSI based on the received LTF (§\ref{sec:backgr}).
In remote attacks, the adversary manipulates this Wi-Fi LTF at a router by adding ``perturbation signals'' to obfuscate HAR at the HAR device, as illustrated in Figure~\ref{fig:sysModel}. The core problem is how to design these perturbation signals, which poses two fundamental challenges:

\circledtextset{resize=real}\circledtext*[height=1.9ex,charshrink=0.65 ]{1} \textbf{Practical remote adversarial attacker systems must be able to operate without any information on HAR devices}.
Unlike digital attacks, the adversary in remote attacks does not have access to any (input) information on HAR devices. For example, exact HAR input CSI and models are unavailable to the adversary; it must generate perturbation signals only using recent CSI data available at the moment of LTF manipulation.
This \emph{mismatch in inputs} between the adversary and HAR makes the design of effective perturbation signals more challenging. A recent remote attack system called C\&W~\cite{li2024practical} has addressed the model mismatch by using a surrogate model, but it still assumes the knowledge of HAR's sliding-window (\emph{i.e.,} input sequence) parameters, such as \emph{HAR time duration}, \emph{sampling rate}, and \emph{input CSI length}, for synchronizing the perturbation with the HAR input window. 
Typical HAR systems, however, employ varying sequence parameters
across implementations~\cite{yousefi2017survey, chen2018wifi, islam2022stc}, and these are unavailable to the adversary.
Thus, such synchronized attacks 
are often impractical.
Table~\ref{tab:relworks} summarizes the limitations of existing Wi-Fi-HAR adversarial systems. 
\circledtextset{resize=real}\circledtext*[height=1.9ex,charshrink=0.65 ]{2} \textbf{Perturbation signals in remote attacks must not cause significant degradation to regular Wi-Fi data communications, the primary function of Wi-Fi systems}. 
In remote attacks, while perturbation signals are added only to the LTF, not the payload, a strong perturbation could lead to severely distorted channels and thus degrade regular communication performance. Therefore, a sophisticated design and power control of perturbation signals are essential. In this regard, multi-antenna diversity, common in Wi-Fi, can help preserve data communication even under perturbation~\cite{marti2023universal}.
In remote attacks, however, multi-antenna systems, such as Single-Input or Multi-Input Multi-Output (SIMO or MIMO) environments, impose a fundamental constraint: LTF manipulation at the router simultaneously affects CSI estimation across multiple antennas at the user, constraining the achievable distortion patterns~\cite{zhou2022wiadv}. As a result, existing remote attack systems, designed for single-antenna settings, cannot leverage multi-antenna gains, as shown in Table~\ref{tab:relworks}.

\parabreak{}In this paper, we present \systemname~(\S\ref{sec:systobj}), a \emph{zero-knowledge} remote adversarial attack system against Wi-Fi HAR, which operates without any prior information about target HAR systems. To do so (challenge \circledtextset{resize=real}\circledtext*[height=1.9ex,charshrink=0.65 ]{1}), 
\systemname{} adopts a surrogate model approach and \emph{generative adversarial imitation learning} (GAIL). 
A surrogate model is an accessible approximation of an unknown target model.
In \systemname{}, the surrogate model is a locally-built Bi-LSTM model, commonly used for Wi-Fi-based HAR classifiers~\cite{guo2019wiar, yousefi2017survey, chen2018wifi, islam2022stc, jannat2023efficient, zhang2020data, khan2020differential}. \systemname{} generates \emph{full-window reference} perturbation signals targeting this surrogate, based on historical CSI and FGM-generated adversarial examples. 
This CSI-to-perturbation mapping allows for generating reference perturbation signals 
without requiring direct access to actual target HAR models, and hence, addresses the model mismatch problem inherent in remote Wi-Fi HAR attacks.
The underlying principle relies on ``adversarial transferability'', where the adversarial examples generated against surrogate models effectively transfer to unknown target systems~\cite{papernot2016transferability}. C\&W directly uses this full-window reference perturbation for LTF manipulation. However, this surrogate model approach alone cannot fully resolve the input mismatch; the full-window reference must synchronize with the HAR's unknown input window, otherwise causing misaligned perturbations (\S\ref{subsec:att_results}).  

\begin{table}
\begin{tiny}
\caption{Comparison of adversarial attacker systems against Wi-Fi HAR systems (see \S \ref{sec:rel_work} for details).}
\vspace{-0.3cm}
\label{tab:relworks}
\centering
\setlength{\tabcolsep}{2.5pt}
\begin{tabularx}{\columnwidth}{@{}
  Y{1.25}| 
  Y{0.75}  
  Y{0.85}  
  Y{1.30}  
  Y{1.00}  
  Y{0.90}  
  Y{0.95}  
@{}}
  \toprule
  \multirow{3}{*}{\begin{tabular}[c]{@{}c@{}}\textbf{\scriptsize{Attacker}}\\\textbf{\scriptsize{system}}\end{tabular}} & 
  \multirow{3}{*}{\begin{tabular}[c]{@{}c@{}}\textbf{Attack}\\\textbf{type}\end{tabular}} & 
  \multirow{3}{*}{\begin{tabular}[c]{@{}c@{}}\textbf{S/MIMO}\\\textbf{compat.}\end{tabular}} & 
  \multicolumn{4}{c}{\textbf{Need for information on target HAR system}} \\
  \cline{4-7}
  & & & 
  \begin{tabular}[c]{@{}c@{}}\textbf{Classifier}\\\textbf{model}\end{tabular} & 
  \begin{tabular}[c]{@{}c@{}}\textbf{HAR}\\\textbf{CSI window}\end{tabular} & 
  \begin{tabular}[c]{@{}c@{}}\textbf{HAR}\\\textbf{duration}\end{tabular} & 
  \begin{tabular}[c]{@{}c@{}}\textbf{Sampling}\\\textbf{rate}\end{tabular} \\
  \midrule
  \AAEN~\cite{zhang2020understanding} & {\cellcolor{LightRed}Digital} & {N/A} & {\cellcolor{LightRed}required} & {\cellcolor{LightRed}required} & {\cellcolor{LightRed}required} & {\cellcolor{LightRed}required}\\
  \sf{ADG}~\cite{zhou2019adversarial} & {\cellcolor{LightRed}Digital} & {N/A} & {\cellcolor{LightRed}required} & {\cellcolor{LightRed}required} & {\cellcolor{LightRed}required} & {\cellcolor{LightRed}required}\\
  \WiCAM~\cite{xu2022wicam} & {\cellcolor{LightRed}Digital} & {N/A} & {\cellcolor{LightRed}required} & {\cellcolor{LightRed}required} & {\cellcolor{LightRed}required} & {\cellcolor{LightRed}required}\\
  Universal~\cite{xie2023universal} & {\cellcolor{LightRed}Digital} & {N/A} & {\cellcolor{LightRed}required} & {\cellcolor{LightRed}required} & {\cellcolor{LightRed}required} & {\cellcolor{LightRed}required}\\
    \ISWARS~\cite{huang2021wars} & {\cellcolor{LightGreen}Remote} & {\cellcolor{LightRed}X} & {\cellcolor{LightRed}required} & {\cellcolor{LightRed}required} & {\cellcolor{LightRed}required} & {\cellcolor{LightRed}required} \\
  \CW~\cite{li2024practical} & {\cellcolor{LightGreen}Remote} & {\cellcolor{LightRed}X} & {\cellcolor{LightGreen}not req.} & {\cellcolor{LightRed}required} & {\cellcolor{LightRed}required} & {\cellcolor{LightRed}required}\\
  \textbf{\textsf{GRAW}} (ours) & {\cellcolor{LightGreen}Remote} & {\cellcolor{LightGreen} \checkmark} & {\cellcolor{LightGreen}not req.} & {\cellcolor{LightGreen}not req.} & {\cellcolor{LightGreen} not req.} & {\cellcolor{LightGreen} not req.}\\
  \bottomrule
\end{tabularx}
\end{tiny}
\vspace{-0.6cm}
\end{table}

\systemname{} enables the zero-knowledge operation by employing GAIL, an \emph{imitation learning} (IL) technique. In \systemname{}, GAIL learns the \emph{temporal relationship} between the surrogate model's input CSI and the generated reference perturbation signal. Once trained, GAIL generates \systemname{}'s perturbation using only recent CSI via online inference, removing the need for synchronized attacks. Like reinforcement learning (RL), GAIL learns a policy that maps states (recent CSI) to actions (online perturbation). However, RL requires reward feedback, \emph{i.e.,} the degraded HAR accuracy in our scenario, which is unavailable during online inference~\cite{williams1992simple}. 
IL resolves this by training the policy to imitate the experts' state-action pairs made by the surrogate (historical CSI and pre-computed reference adversarial examples), without immediate reward feedback. Among IL methods, behavioral cloning, which utilizes conventional time-series neural networks, can also learn this temporal mapping, but its performance often degrades in unseen environments due to covariate shift~\cite{gong2019real}. GAIL mitigates this issue by learning a policy that adapts to changes in the environment through adversarial training~\cite{ho2016generative}. To our best knowledge, \systemname{} is the first Wi-Fi HAR adversarial system to leverage GAIL, thus completely eliminating the need for information on target HAR systems.

\systemname{} also minimizes the degradation in Wi-Fi communication performance (challenge \circledtextset{resize=real}\circledtext*[height=1.9ex,charshrink=0.65 ]{2}) by dynamically controlling the perturbation signal amplitudes.
Specifically, \systemname{} 
adjusts each perturbation signal power based on the running average of past signals, keeping its ratio to the LTF amplitude under a target threshold (\S\ref{sec:ampAdjust}). 
In addition, as \systemname{} determines each online perturbation from the recent samples, its power distribution is flatter than the reference perturbations, preventing spikes in the perturbation signal powers, which are critical to communication degradation. \systemname{} also addresses the aforementioned S/MIMO constraint by projecting the multi-antenna perturbation signals onto a shared single LTF, to best approximate the intended CSI distortion across all receive antennas (\S\ref{sec:advMIMO}). This S/MIMO compatibility allows for keeping improving the regular data communications performance with more antennas using spatial diversity (\emph{e.g.,} Maximum Ratio Combining).




We evaluate \systemname{} against seven target architectures using datasets collected in five environments, including our own dataset. Our evaluation encompasses scenarios of unknown input timing and sampling rate as well as unmatched models between surrogate and HAR classifiers. From our evaluations, it is observed that \systemname{} is the only remote attack scheme capable of degrading every evaluated HAR model to a random-selection level, with up to 76.7\% higher \emph{attack success ratio} (ASR) than existing methods at the same perturbation amplitude. 
We also demonstrate the feasibility of \systemname{} through real-time, over-the-air (OTA) experiments using software-defined radios (SDR). We implement \systemname{}'s pipeline — the inference model, MIMO processing, and online amplitude adjustment — as real-time C++ GNU Radio blocks for $2\times2$ MIMO. The hardware demonstration shows that \systemname{} with 1000-byte packets achieves over 50\% ASR at \emph{perturbation-to-signal ratio} (PSR) -4\,dB while maintaining a 99.7\% packet success rate under spatial multiplexing.


The source code of \systemname{} is publicly available.\footnote{https://github.com/byungjunkim12/25-adv\_HAR}


\section{Background}
\label{sec:backgr}

\subsection{Wi-Fi-based HAR}
In Wi-Fi communications, the orthogonal frequency-division multiplexing (OFDM)-MIMO system with a transmitter (TX) with $\nTX$ antennas and a receiver (RX) with $\nRX$ antennas is modeled as:
\begin{equation}
\textbf{y}_{ij} = \textbf{H}_{ij}\textbf{x}_{ij} + \textbf{n}_{ij}
\end{equation}
with $1\le i \le M$, $1 \le j \le \nSC$, $\textbf{x}_{ij} \in \mathbb{C}^{\nTX}$, $\textbf{y}_{ij}\in \mathbb{C}^{\nRX}$, $\textbf{H}_{ij}\in \mathbb{C}^{\nRX\times \nTX}$, and $\textbf{n}_{ij} \in \mathbb{C}^{\nRX}$ denoting the transmitted signal, the received signal, the CSI matrix, and the noise vector respectively. An RX uses known LTF ($\textbf{x}_{ij}$) for channel estimation ($\textbf{H}_{ij}$) using the received signals ($\textbf{y}_{ij}$).

In our HAR scenarios, a Wi-Fi router is a TX, while an HAR device is an RX. An HAR classifier at the user device typically takes a sequence of CSI matrices from TX as input and outputs the probability of each activity. LSTM is commonly used for a neural network model of the classifier in Wi-Fi HAR~\cite{yousefi2017survey, chen2018wifi, sheng2020deep, ding2020rf}. LSTM HAR classifier, $f_C$, also takes as an input sequence of CSI matrices, $\HHH \in \mathbb{R}^{M\times\nSC \times \nRX \times \nTX}\triangleq \{|\HH_{ij}|\}_{1\le i\le M,1\le j\le \nSC}$, where $|\HH_{ij}|$ is the matrix containing the amplitudes of $\HH_{ij}$'s elements. The classifier outputs a vector of elements, each of which represents the probability of each activity. Since different activities take different durations, the length of each input sequence is not fixed. Since activity durations vary, \systemname{} adopts a Bi-LSTM-based surrogate model (\S\ref{sec:perturb}), which handles variable-length CSI inputs.

\subsection{Adversarial Attacks}
Adversarial attacks against neural networks have been widely studied, as carefully crafted perturbations can significantly degrade the performance of a target model even when they are (nearly) imperceptible. In our HAR attack setting, the classifier inputs are CSI sequences, so the perturbed CSI serves as the \emph{adversarial example}. Thus, the adversarial example, a perturbed input, is 
$\hat{\HHH} = \HHH + \hat{\AAA}$
that remains close to the original input \(\HHH\) in some norm, but is intentionally designed so that the classifier’s prediction changes (e.g., \(f(\hat{\HHH}) \neq f(\HHH)\)). 
For example, FGM~\cite{goodfellow2014explaining} creates an adversarial example $\hat{\HHH}$:
\begin{equation}
\hat{\HHH} = \HHH + \alpha \nabla_\HHH{\mathcal{L}\left(f(\HHH), \textbf{z}\right)}
\label{eq:FGM}
\end{equation}
where 
\(\mathbf{z} \in \mathbb{R}^{N_C}\) is the one-hot encoded label of \(\mathbf{\HHH}\), \(\alpha\) is a parameter to control the perturbation magnitude, \(\mathcal{L}\) is the loss function of \(f\), and $N_C$ is the number of classes.

FGM alone, however, cannot compute adversarial examples when the adversary lacks access to the target classifier information. This scenario, known as a \textit{black-box attack}~\cite{papernot2016transferability}, occurs when the adversary has limited knowledge of the target model, such as its architecture or training data. In this case, one can employ a \emph{surrogate model}, $\fCsur(\HHH)$, instead of $f(\HHH)$, to generate adversarial examples using only adversary-accessible data. This approach relies on \emph{adversarial transferability}~\cite{papernot2016transferability}, where adversarial examples made against a surrogate model remain effective against unknown target models. In this work, we specifically focus on the case where the adversary has ``no information'' about the target model, except for the ground-truth activity labels of the available training data. 

\begin{figure*}
\centering
\includegraphics[width=\textwidth]{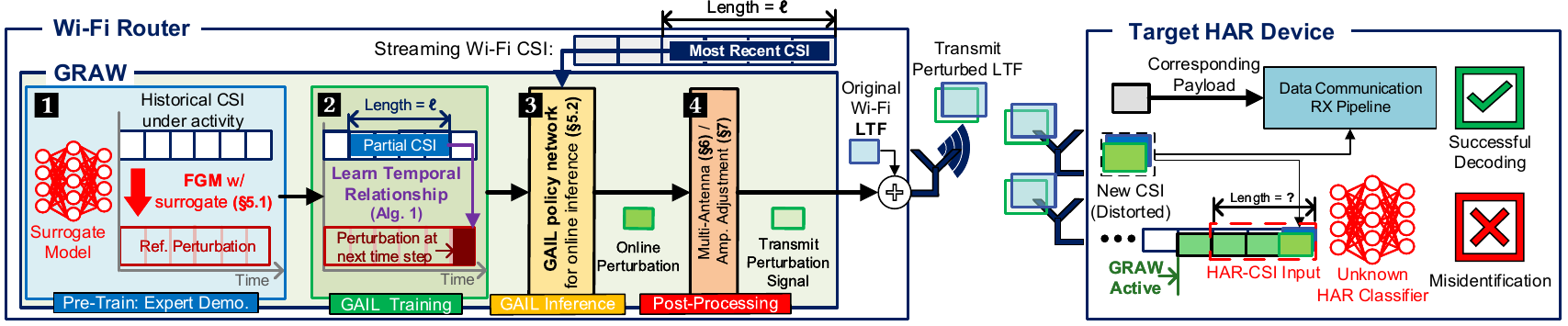}\vspace{-0.35cm}
\caption{\systemname{}'s remote adversarial attack operation.}\label{fig:GRAW_oper}
\end{figure*}

\subsection{Reinforcement and Imitation Learning}
\label{subsec:IL}
\systemname{} utilizes a Reinforcement Learning (RL) architecture to address a real-time adversarial attack problem. In real-time adversarial attack scenarios, adversarial examples must be computed at each time step using the sequence of CSI estimated up to that point. This challenge aligns with the RL paradigm, where a policy function $\pi(a_i|s_i)$ determines actions $a_i$ based on observed states $s_i$ and influences subsequent state transitions. RL aims to find the optimal policy function $\pi$ that maximizes the cumulative reward function over a trajectory, ${\tau=\{(s_i, a_i)}\}_{i=1}^M$.

In our Wi-Fi HAR attack scenarios, estimated CSI up to the current time, paired with corresponding adversarial examples, form a state-action pair and the reward function represents the degraded accuracy of the target HAR. In RL, such as REINFORCE~\cite{williams1992simple}, the learning agent interacts with the reward function for feedback on its actions. However, this interaction is infeasible in our attack scenarios, since the adversary determines the accuracy of the target classifier only after processing the entire input. Thus, \systemname{} resolves this issue by adopting \emph{Imitation Learning} (IL), a specialized form of RL, where a training agent learns to replicate expert behavior solely from given expert trajectories (\S\ref{subsec:GAIL}). 

\vspace{-5pt}
\section{Related Work and Limitation}
\label{sec:rel_work}
Adversarial attacks targeting Wi-Fi-based HAR systems for privacy protection have attracted significant research attention; we summarize and compare prior works in Table~\ref{tab:relworks}.

Early studies focused on digital attacks, where techniques from other adversarial ML domains, such as image classification, could be more directly applied since the attacker directly manipulates the classifier input. Early works on digital attacks, such as \AAEN~\cite{zhou2019adversarial} and {\sf{ADG}}~\cite{zhang2020understanding}, suggest modifying the HAR classifier's loss functions to prevent the detection of specific activities. However, these approaches assume adversaries can retrain classifiers—a highly unrealistic threat model in practice, considering that adversaries need to interfere with the training process. Other approaches, like {\sf{WiCAM}}~\cite{xu2022wicam} and Ref.~\cite{xie2023universal}, manipulate the CSI input sequences rather than attacking during training, eliminating model training access requirements. Nevertheless, these methods still require direct access to the target HAR system input sequences, which is impractical since the CSI data in user devices is not publicly accessible.

To address the impracticality of digital attacks, remote attack methods have emerged where adversaries operate outside target devices without direct system access. A denial-of-service approach~\cite{liu2023time} jams packet transmissions via Wi-Fi collision-avoidance protocols, preventing HAR systems from receiving CSI sequences but also disrupting normal data communication. {\sf PhyCloak}~\cite{qiao2016phycloak} instead deploys a full-duplex relay that selectively obfuscates RF features for illegitimate Wi-Fi sensing while preserving a designated legitimate sensor. However, it targets traditional RF sensors that extract physical features such as Doppler shifts, and its selective preservation depends on physically co-locating the legitimate sensor with the relay, requiring dedicated hardware. Its single-antenna obfuscation further limits it to SISO settings. Other works, including WiAdv~\cite{zhou2022wiadv}, IS-WARS~\cite{huang2021wars}, LTF-based \CW{} perturbations~\cite{li2024practical}, and LTF-based FGM~\cite{kim2025real}, build over-the-air adversarial or spoofing signals, but require unrealistic assumptions, such as the knowledge of target model architectures and perfect synchronizations with the target input. Furthermore, most of the remote attack systems assume single-antenna scenarios, as manipulating LTF signals from a TX 
affects CSI estimation across multiple RX antennas simultaneously, constraining the achievable distortion patterns. In contrast, \systemname{} generates perturbation signals without requiring knowledge of target models or synchronization requirements, while explicitly addressing practical multi-antenna RX constraints.

\vspace{-5pt}
\section{\systemname{} Overview}
\label{sec:systobj}


\systemname{} is a privacy-protecting adversary at the router against the HAR device's activity recognition. 
\systemname{} modifies the downlink LTF so that the CSI estimated at the HAR device becomes an adversarial example to the HAR classifier, causing misclassification. \systemname{} achieves this 
without significantly degrading Wi-Fi data communications (\S\ref{s:impact_link}, \S\ref{subsec:real-time}).


\systemname{} requires no prior knowledge on target HAR devices; we refer to this a \textbf{zero-knowledge operation} (\emph{i.e.,} zero knowledge on the target) in this paper. 
It relies only on a surrogate classifier trained on CSI pre-collected for the target activities. For deployments, we envision that 
the users (de)activate \systemname{} on demand to protect their privacy, sharing their activity timing; we also discuss its automated opportunistic operation in Section~\ref{sec:disc}.

\parahead{Operation.} 
Figure~\ref{fig:GRAW_oper} overviews \systemname{}'s operation. For remote attacks, \systemname{} computes and adds a \emph{transmit perturbation signal}
to the router's LTF, through the following stages:

\parahead{\circledtext*[boxtype=O,height=2ex,charshrink=0.7 ]{1} Expert Demonstration Generation.} \systemname{} first builds the surrogate HAR classifier using historical CSI under activity and generates \emph{reference perturbations} by applying FGM to this surrogate (\S\ref{subsec:blackFGM}). This provides the expert demo to GAIL.

\parahead{\circledtext*[boxtype=O,height=2ex,charshrink=0.7 ]{2} GAIL Training.} Based on this demo, the GAIL-based \emph{online perturbation} generator is trained to learn the temporal relationship between the $\ell$ samples from the historical CSI and the corresponding step reference perturbation (\S\ref{subsec:GAIL}).

\parahead{\circledtext*[boxtype=O,height=2ex,charshrink=0.7 ]{3} GAIL Inference.} The perturbation generator computes online perturbation $\AA_i\in \mathbb{R}^{\nSC\times\nRX\times\nTX}$, at time step $i$, based on the $\ell$ most recent CSI samples estimated at the router.

\parahead{\circledtext*[boxtype=O,height=2ex,charshrink=0.7 ]{4} Post-processing.} When the RX has multiple antennas ($\nRX > 1$), $\AA_i$ cannot be applied directly, as a single LTF is shared across all RX antennas. \systemname{} therefore converts $\AA_i$ into $\bar{\BB}_i \in \mathbb{R}^{\nSC\times\nTX}$ whose combined effect at the RX best approximates $\AA_i$ (\S\ref{sec:advMIMO}). Finally, the amplitudes of $\bar{\BB}_i$ are adjusted online to keep its ratio to the LTF under a target threshold, yielding the transmit perturbation signal $\BB_i$ (\S\ref{sec:ampAdjust}). 

\systemname{} then adds transmit perturbation signals $\BBB \triangleq \{\BB_i\}_{i=1}^M \in \mathbb{R}^{M\times\nSC \times \nTX}$ to LTF. Note that only Step \circledtext*[boxtype=O,height=2ex,charshrink=0.7 ]{3} \& \circledtext*[boxtype=O,height=2ex,charshrink=0.7 ]{4} are online.
At the RX, the CSI matrix is estimated using the perturbed LTF, which leads to degradation of the target HAR classifier $\fC$. Thus, \systemname{} seeks $\BBB$ that minimizes the accuracy of $\fC$. 


\systemname{} requires two capabilities: extracting CSI and manipulating the transmitted LTF. CSI extraction is standard on commodity NICs (up to 1~kHz~\cite{halperin2011tool}, far above \systemname{}'s up to 50~Hz update rate). LTF manipulation, however, requires physical-layer access beyond commodity NICs; we realize it with an SDR (\S\ref{subsec:real-time}) and discuss commodity deployment in (\S\ref{sec:disc}). By Wi-Fi's TDD channel reciprocity, the CSI the router estimates from a received packet equals the CSI the HAR device observes, provided that a packet is exchanged within the channel coherence time.
\systemname{} therefore updates the perturbation within the channel coherence time, 20~ms for 5~GHz Wi-Fi under human movement at 1.5~m/s ($f_d = 25$~Hz, $T_c = 1/(2f_d) \approx 20$~ms)~\cite{Xie2015Precise}, which also satisfies the 41.7~ms coherence time at 2.4~GHz.
\vspace{-5pt}
\section{Online Perturbation Generation}
\label{sec:perturb}

\systemname{} trains its online perturbation generator via GAIL, as illustrated in Figure~\ref{fig:GRAW_oper}. First, the adversary builds a surrogate model to compute \emph{reference} perturbation (\S\ref{subsec:blackFGM}). Using this surrogate model, black-box FGM computes these reference perturbations on the surrogate, which are then paired with the CSI the router estimates. Using GAIL, the policy network is then trained to map the CSI sequences to online perturbations by imitating references. (\S\ref{subsec:GAIL}).

\subsection{Surrogate Model}
\label{subsec:blackFGM}

The goal of the surrogate model is to compute reference perturbations that will serve as expert demonstrations for training the GAIL policy network later.

Since the target HAR classifier is unknown to the adversary, we train a surrogate model using available CSI data. We adopt a Bi-LSTM layer as the surrogate classifier $\fCsur$, commonly used in Wi-Fi-based HAR classifiers~\cite{guo2019wiar, yousefi2017survey, chen2018wifi, islam2022stc, jannat2023efficient, zhang2020data, khan2020differential}. Owing to the transferability of adversarial examples across deep learning models~\cite{papernot2016transferability}, perturbations computed against the surrogate are expected to be effective against the target HAR classifier as well, and we therefore use them as expert demonstrations for the GAIL policy network. Using the surrogate, black-box FGM computes $\hat{\AAA} \in \mathbb{R}^{M\times\nSC \times \nRX \times \nTX}$ and thus an adversarial example:
\begin{equation}
\hat{\HHH} = \HHH + \alpha \nabla_\HHH{\mathcal{L}\left(\fCsur(\HHH), \textbf{z}\right)} = \HHH + \hat{\AAA},
\label{eq:blackFGM}
\end{equation}
where $f$ in~\eqref{eq:FGM} is replaced with the surrogate model $\fCsur$. The resulting $\hat{\AAA}$ then feeds into the GAIL policy network as expert demonstrations (\S\ref{subsec:GAIL}). We also present the performance of black-box FGM as a reference baseline (\S\ref{subsec:att_results}), although it cannot be directly used for remote attacks against Wi-Fi HAR since it requires the CSI sequence over entire action, including those not yet observed at attack time.

\subsection{GAIL-based Perturbation Generator}
\label{subsec:GAIL}
In a remote attack scenario, perturbation signals must be generated online, as the HAR input CSI may be unavailable at generation time. To address this, our GAIL-based perturbation generator is trained to imitate the mapping from recent CSI to the reference perturbation (\S\ref{subsec:blackFGM}) computed from the surrogate via FGM at the current time step.

The GAIL-based generator takes the $\ell$ most recent CSI estimates preceding time step $i$, $\HHH_i^{\ell} \triangleq \{\HH_j\}_{j = i-\ell}^{i-1}\in \mathbb{R}^{\ell \times\nSC \times \nRX \times \nTX}$, and outputs an online perturbation, $\AA_i$. The goal is to find the policy function $\pi(\AA_i|\HHH_i^{\ell})$ that minimizes the accuracy of \(\fC(\HHH+\AAA)\), where \(\AAA \triangleq \{\AA_i\}_{i=1}^M\sim \pi(\cdot|\HHH_i^{\ell})\). Since the HAR classifier $\fC$ takes the entire CSI sequence $\HHH$ as input, directly optimizing this objective requires the entire sequence that is unavailable at time step $i$. We instead train $\pi$ to imitate the reference perturbations $\hat{\AAA}$ computed by black-box FGM in Eq.~\eqref{eq:blackFGM}. The generator thereby learns to reproduce these references from only the past CSI $\HHH_i^{\ell}$.

To learn this imitation, we employ GAIL, an IL algorithm that learns from expert trajectories, $\{\HHH_i^{\ell}, \hat{\AA}_i\}_{i=1}^M$. We choose GAIL because it generalizes to unseen environments, unlike behavioral cloning, which does not~\cite{ho2016generative} (\S\ref{subsec:att_results}). GAIL casts imitation as adversarial distribution matching: like a generative adversarial network (GAN), it comprises a discriminator ($D_w$) that distinguishes expert from learner trajectories $\{\HHH_i^{\ell}, \AA_i\}_{i=1}^M|_{\AA_i\sim \pi_\theta(\cdot|\HHH_i^{\ell})}$ and a policy ($\pi_\theta$) trained to fool it. The complete GAIL objective is:
\begin{equation}
\begin{aligned}
    &\min_{\pi_\theta}{\max_{D_w}~} {\mathbb{E}_{(\HHH_i^{\ell},\AA_i)\sim\pi_{\theta}(\AA_i|\HHH_i^{\ell})}[\log(D_w(\HHH_i^{\ell},\AA_i))]+} \\
    &\mathbb{E}_{(\HHH_i^{\ell},\hat{\AA}_i)}[\log(1-D_w(\HHH_i^{\ell},\hat{\AA}_i)]-\lambda_G H(\pi).
\end{aligned}
\end{equation}
Detailed optimization steps are presented in Algorithm~\ref{alg:GAILattack}.

\begin{algorithm}
\caption{Training the online perturbation generator using GAIL}\label{alg:GAILattack}
\KwData{Expert trajectories $\tau_E=\{\HHH_i^{\ell}, \hat{\AA}_i\}$ where $i=\{1,2,\cdots,M\}$, initial parameters for discriminator $w_0$ and policy function $\theta_0$}
\For{$k=0,1,\cdots,K-1$}{
Sample trajectories using the learner policy $\tau_k\sim\pi_{\theta_k}(\AA_i|\HHH_i^{\ell})$\;
Update discriminator parameters to increase the objective: $w_{k+1}\leftarrow w_k + \nabla_{w_k}J(w_k)$~\eqref{eq:GAILdis}
\\
Update policy function parameters to decrease the objective: $\theta_{k+1}\leftarrow \theta_k - \nabla_{\theta_k}K(\theta_k)$~\eqref{eq:GAILpol}
}
\KwOut{Trained policy network $\pi_{\theta_K}(\AA_i|\HHH_i^{\ell})$ that generates online perturbations}
\end{algorithm}

For each iteration of Algorithm~\ref{alg:GAILattack}, the learner trajectories $\tau=\{\HHH_i^{\ell}, \AA_i\}$ pair CSI available to the router with the corresponding online perturbations $\AA_i$ that the policy function $\pi_{\theta_k}(\cdot|\HHH_i^{\ell})$ generates. The discriminator and policy functions are alternately optimized using these learner and expert trajectories. In line 3, the discriminator function parameters, $w$, are updated using the gradient:
\begin{equation}
\label{eq:GAILdis}
\begin{aligned}
\nabla_wJ(w) = \mathbb{E}_{(\HHH_i^{\ell},\AA_i)\sim\pi_{\theta_k}}[\nabla_{w}\log (D_w(\HHH_i^{\ell},\AA_i))] \\
+\mathbb{E}_{(\HHH_i^{\ell},\hat{\AA}_i)}[\nabla_w\log (1-D_w(\HHH_i^{\ell},\hat{\AA}_i))].
\end{aligned}
\end{equation}
Line 4 describes the policy gradient, which minimizes the cost function evaluated on learner trajectories while maximizing the regularizer to encourage exploration,
\begin{small}
\begin{equation}
\label{eq:GAILpol}
\nabla_{\theta}K(\theta)=~\mathbb{E}_{\tau_i}[\nabla_\theta\log \pi_{\theta}(\AA_i|\HHH_i^{\ell})C(\HHH_i^{\ell},\AA_i)] - \lambda_G\nabla_{\theta} H(\pi_{\theta}),
\end{equation}
\end{small}
where cost function $C(\HHH_i^{\ell}, \AA_i) = \mathbb{E}_{\tau_k}[\log(D_{w_{k+1}}(\HHH_i^{\ell}, \AA_i))]$. As $C(\HHH_i^{\ell}, \AA_i)$ depends on policy through the sampled trajectory $\tau_k$, computing $\nabla_{\theta}K(\theta)$ is non-trivial. As in GAIL, we use trust region policy optimization (TRPO)~\cite{schulman2015trust} (Appendix~\ref{app:TRPO}).
\section{Multi-Antenna LTF Manipulation}
\label{sec:advMIMO}

When an HAR device has multiple antennas (\emph{i.e.,} $\nRX$ > 1), manipulating a single LTF from a router affects CSI estimation at all receiver antennas, making it impossible to arbitrarily adjust multiple CSIs through a single LTF modification. However, the online perturbation $\AAA_i$ from GAIL (\S\ref{subsec:GAIL}) is designed to adjust each receiver antenna's CSI independently, creating dimension mismatch with the LTF symbols; thus it cannot be added directly. To overcome this limitation, we design a multi-antenna LTF manipulation scheme based on projection, which generates the transmit perturbation signal $\BB_i$ by projecting $\AA_i$ onto a one-dimensional subspace spanned by the SIMO channel vector. This projection finds the LTF perturbation whose resulting CSI estimate is as close as possible to the target, i.e., the sum of $\HH_i$ and $\AA_i$.

For the $k$-th antenna of the TX, the adversary adds perturbation, $\textit{b}_{ij}^k \triangleq[\BB_i]_{jk} \in \mathbb{R}$, to LTF, $x_{ij}^k\in \mathbb{R}$, at $i$-th time step, $j$-th subcarrier. The goal is to make the router misestimate the original CSI, $\textbf{\textit{h}}_{ij}^k\in \mathbb{R}^{\nRX}$ as $\textbf{\textit{h}}_{ij}^k + \textbf{a}_{ij}^k$, where $\textbf{a}_{ij}^k\in \mathbb{R}^{\nRX}$ is the corresponding element of the output of GAIL, $\AAA$. To achieve this goal, the perturbation must satisfy:
\begin{equation}
\begin{aligned}
\textbf{\textit{h}}_{ij}^k (x_{ij}^k + \textit{b}_{ij}^k) / x_{ij}^k & = (\textbf{\textit{h}}_{ij}^k + \textbf{a}_{ij}^k) \Rightarrow
\textbf{a}_{ij}^k = \textbf{\textit{h}}_{ij}^k \textit{b}_{ij}^k /x_{ij}^k.
\end{aligned}
\end{equation}

However, when $\nRX>1$, an exact solution for $\textit{b}_{ij}^k$ may not exist. Thus, we project the online perturbation $\textbf{a}_{ij}^k$ onto the feasible space, yielding $\bar{\textit{b}}_{ij}^k$:
\begin{equation}
\label{eq:advMIMO}
\bar{\textit{b}}_{ij}^k=\arg\min_{\textit{b}_{ij}^k}||\textbf{\textit{h}}_{ij}^k \textit{b}_{ij}^k /x_{ij}^k - \textbf{a}_{ij}^k|| = x_{ij}^k (\textbf{\textit{h}}_{ij}^k \cdot \textbf{a}_{ij}^k) / ||\textbf{\textit{h}}_{ij}^k||^2.
\end{equation}
$\bar{\textit{b}}_{ij}^k$ is then added to the original LTF, and the router antenna $k$ transmits the manipulated LTF $x_{ij}^k+\bar{\textit{b}}_{ij}^k$.
We experimentally validate this in \S\ref{subsec:validation}, where GRAW achieves comparable performance to the ideal case (LTF-oracle) where a separate LTF is used for each TX-RX antenna pair. Since each TX antenna transmits an orthogonal LTF whose CSI is estimated separately at the RX, the projection in~\eqref{eq:advMIMO} applies independently per TX antenna, extending directly to MIMO.

\begin{figure}
\centering
    \hfill
        \begin{subfigure}[b]{0.16\columnwidth}
            \hfill
            \includegraphics[width=1\linewidth]{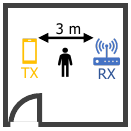}
            \caption{\TAR}
            \label{subfig:env_TAR}
        \end{subfigure}
    \hfill
        \begin{subfigure}[b]{0.48\linewidth}
            \hfill
            \includegraphics[width=1\linewidth]{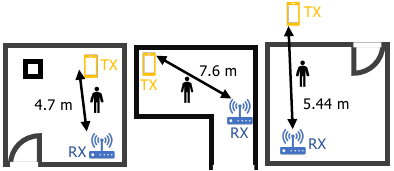}
            \caption{\JAR}
            \label{subfig:env_JAR}
        \end{subfigure}        
    \hfill
        \begin{subfigure}[b]{0.32\linewidth}
            \hfill
            \includegraphics[width=1\linewidth]{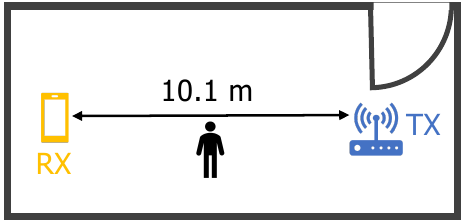}
            \caption{\RUAR}
            \label{subfig:env_RUAR}
        \end{subfigure} 
\vspace*{-0.3cm}\caption{Data collection environments.}
\label{fig:data_env}
\end{figure}

\begin{figure}
\centering
    \hfill
        \begin{subfigure}[t]{0.42\columnwidth}
            \hfill
            \includegraphics[width=1\linewidth]{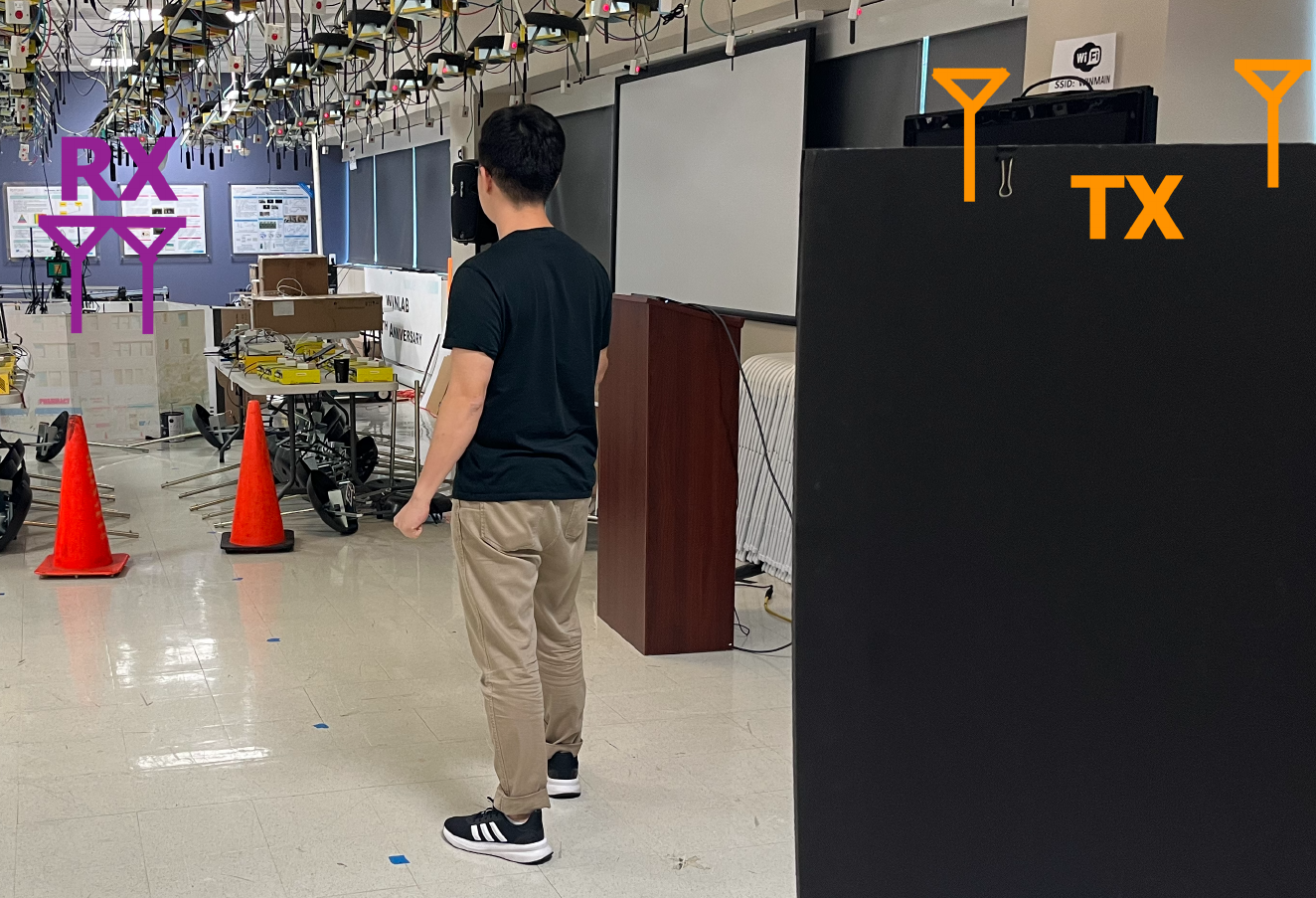}
            \caption{\RUAR{} environment.}
            \label{subfig:RUAR_env}
        \end{subfigure}
    \hfill
        \begin{subfigure}[t]{0.55\columnwidth}
            \hfill
            \includegraphics[width=1\linewidth]{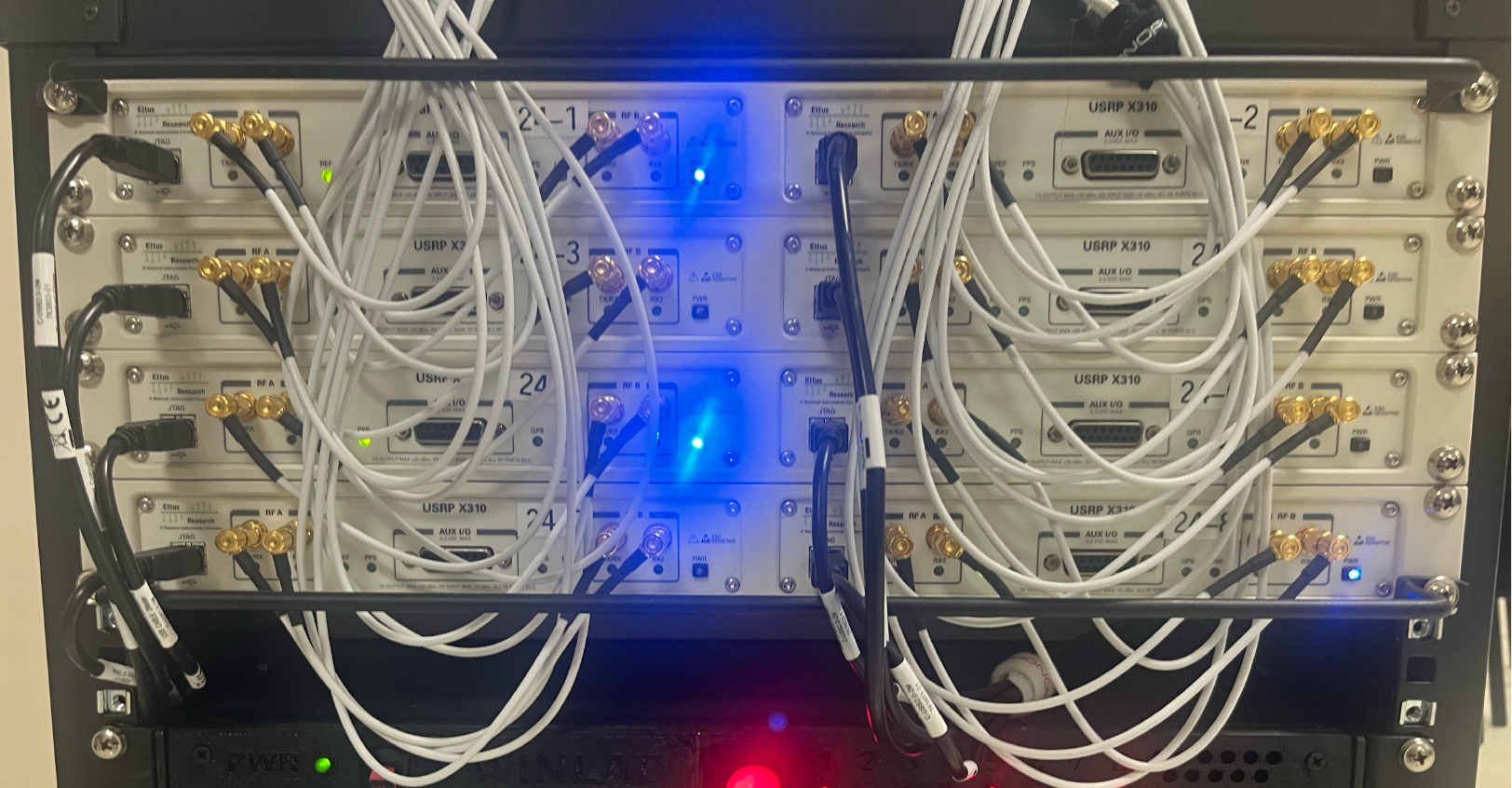}
            \caption{USRP X310 SDR devices.}
            \label{subfig:USRP}
        \end{subfigure}    
\vspace*{-0.3cm}\caption{Our experimental measurement setup (\RUAR{}).}
\label{fig:env}
\end{figure}

\section{Online Amplitude Adjustment}
\label{sec:ampAdjust}
To regulate the impact on the communication link, \systemname{} scales the perturbation signals so their average amplitude ratio to the LTF meets a target $\gamma$, while preserving the temporal power distribution of $\bar{\BB}_i$ from the MIMO processing (\S\ref{sec:advMIMO}). Since the future perturbation signals are unknown during online perturbation generation, we adjust the amplitudes of $\bar{\BB}_i\triangleq\{\bar{\textit{b}}_{ij}^k\}_{j,k}$ in Eq.~\eqref{eq:advMIMO} using only $\bar{\BBB}$ up to the current time step, $\bar{\BBB}_{1:i}=\{\bar{\BB}_{m}\}_{m\le i}$. $\gamma$ is chosen to balance attack strength and communication impact.
\begin{equation}
\BBB_i = \gamma \bar{\BBB}_i \cdot {\|\textbf{x}_{1:i}\|} / {\|\bar{\BBB}_{1:i}\|}.
\label{eq:ampAdjust}
\end{equation}
The rationale behind Eq.~\eqref{eq:ampAdjust} is that $\|\bar{\BBB}_{1:i}\| / \|\textbf{x}_{1:i}\|$ reliably estimates the amplitude ratio over the entire activity. This estimator works effectively because $\bar{\BBB}_{1:i}$ exhibits a flat power distribution over time, unlike the gradient-based attack (further discussed in \S\ref{subsec:att_results}). We also validate this in \S\ref{subsec:validation}.




\section{Evaluation}
\label{sec:eval}
\subsection{Datasets and Target Models}
\label{subsec:data_target}

\begin{figure*}
\centering
    \hfill \begin{subfigure}[t]{0.25\textwidth}
            \hfill \includegraphics[width=1\linewidth]{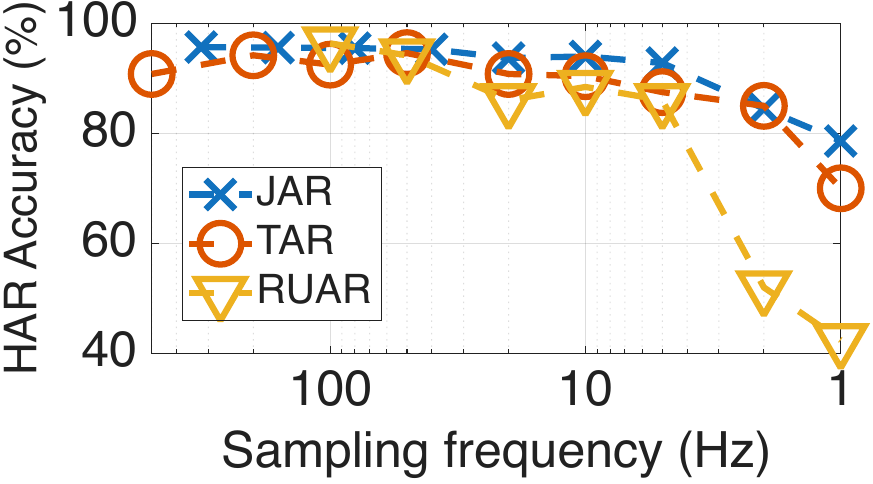}
            \caption{Surrogate model accuracy vs. CSI sampling rate.}
            \label{subfig:surro_dSamp}
        \end{subfigure}
    \hfill \begin{subfigure}[t]{0.73\textwidth}
            \hfill \includegraphics[width=1\linewidth]{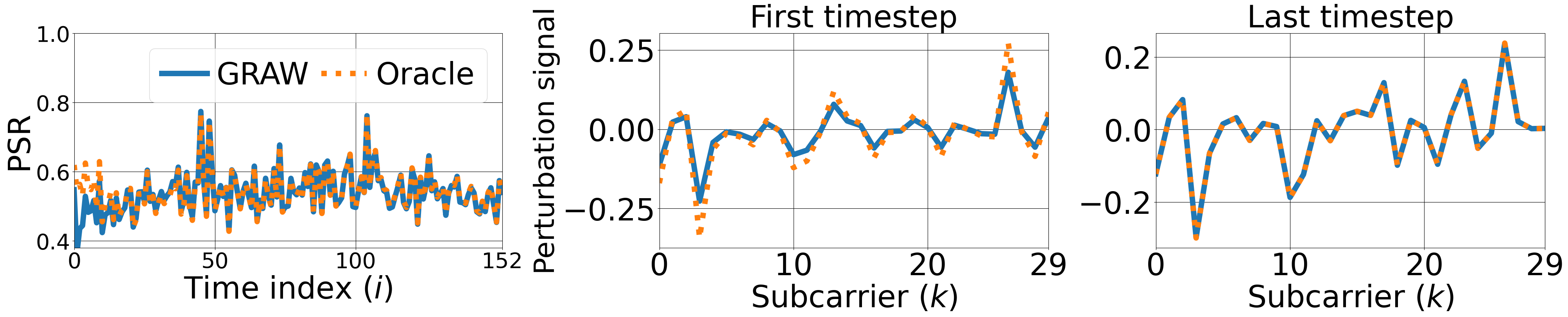}
            \caption{Perturbation signals for one "lie down" sample in the \JAR~dataset (\systemname{} vs. Oracle): $\|\BB_i\|$ vs. $\gamma \cdot \|\bar{\BB}_i\|$ over time (left); $\BB_1$ vs. $\gamma \cdot \bar{\BB}_1$ (middle); and $\BB_M$ vs. $\gamma \cdot \bar{\BB}_M$ (right) across subcarriers.}
            \label{subfig:ampAdjust}
        \end{subfigure} \vspace*{-0.2cm}
\caption{Experimental validation of GRAW’s design components: the surrogate model taking downsampled CSI sequences (we select 50, 40, and 50 Hz on TAR,
JAR, and RUAR, respectively) in Figure~\ref{subfig:surro_dSamp}, while the real-time amplitude adjustment in Figure~\ref{subfig:ampAdjust}. PSR stands for Perturbation-to-Signal Ratio.}
\label{fig:prelim_eval}
\end{figure*}

\newlength{\oldtabcolsep}
\setlength{\oldtabcolsep}{\tabcolsep}
\setlength{\tabcolsep}{2.6pt}
\begin{table}[t]
\begin{scriptsize}
\centering
\caption{Dataset parameters}\vspace*{-0.3cm}
\label{tab:dataParam}
\begin{tabular}{c|ccccccc}
\toprule
\textbf{Dataset} & 
\makecell{\textbf{Carrier}\\\textbf{freq.}\\\textbf{(GHz)}} & 
\makecell{$\{\nTX,$\\$\nRX\}$} & 
\makecell{\textbf{No.}\\\textbf{Activities}} & 
\makecell{\textbf{No.}\\\textbf{Envs.}} & 
\makecell{\textbf{Eval.}\\\textbf{Days}} & 
\makecell{\textbf{No.}\\\textbf{Participants}} & 
\makecell{\textbf{Sampling}\\\textbf{Rate}\\\textbf{(Hz)}} \\ \midrule
\TAR  & 5 & \{1, 3\} & 6 & 1 & 8 & 6  & 1000 \\ \hline
\JAR  & 2.4 & \{1, 3\} & 6 & 3 & 1 & 30 & 320  \\ \hline
\RUAR & 2.4 & \{2, 2\} & 5 & 1 & 3 & 6  & 100  \\ 
\bottomrule
\end{tabular}
\end{scriptsize}
\end{table}
\setlength{\tabcolsep}{\oldtabcolsep} 

\begin{table}[]
\begin{scriptsize}
\centering
\caption{Wi-Fi-based HAR classifiers}\vspace*{-0.3cm}
\centering
\label{tab:HARclass}
\begin{tabular}{c|ccccc}
\toprule
\textbf{HAR model} & \textbf{Dataset} & \makecell{\textbf{Input}\\\textbf{features}}  & \makecell{\textbf{Classifier}\\\textbf{structure}} & \makecell{\textbf{Input}\\\textbf{length}} & \makecell{\textbf{Sampling}\\\textbf{rate (Hz)}} \\ \midrule
Model A~\cite{yousefi2017survey} & \TAR & CSI & LSTM & 2~s & 500 \\ \hline
Model B~\cite{chen2018wifi} & \TAR & CSI & \makecell{LSTM +\\Attention
} & 2~s & 500 \\ \hline
Model C~\cite{islam2022stc} &\JAR & CSI & \makecell{CNN +\\LSTM} & 1.6~s & 1000\\ \hline
Model D~\cite{jannat2023efficient} & \JAR & \makecell{Statistical\\features} & RF & 1.6~s & 1000\\ \hline
Model E~\cite{zhang2023imgfi} & \RUAR & STFT & CNN & Variable & 100\\\hline
Model F~\cite{luo2024vision} & \RUAR & STFT & Transformer & 2~s & 100 \\ \bottomrule
\end{tabular}
\end{scriptsize}
\end{table}

We evaluate \systemname{} using two public datasets, \TAR~\cite{yousefi2017survey} and~\JAR~\cite{baha2020dataset}, as well as a dataset we collect using SDR called \RUAR. These datasets together span both 2.4~GHz (\JAR, \RUAR)  and 5~GHz (\TAR) Wi-Fi frequency bands. Table~\ref{tab:dataParam} summarizes the dataset specifications, and Figure~\ref{fig:data_env} illustrates data collection environments, with floor plans sourced from the original dataset documents~\cite{yousefi2017survey, baha2020dataset}. \JAR~is collected across three distinct environments, including a non-line-of-sight (NLOS) scenario, and \TAR~spans eight days, enabling evaluations under both spatial and temporal diversity.

\parahead{\textbf{RUAR}: Our SDR-based dataset.} \RUAR{} uses a $2\times2$ MIMO setting, while \TAR{} and \JAR{} use $1\times3$ SIMO. Figure~\ref{subfig:env_RUAR} and Figure~\ref{fig:env} describe the environment where \RUAR{} is measured. We deploy two SDRs, USRP X310, for both transmitting and receiving signals. We collect 1,440 activity samples from 6 volunteers over 3 different days.

\begin{figure}
\centering
   \includegraphics[width=\columnwidth]{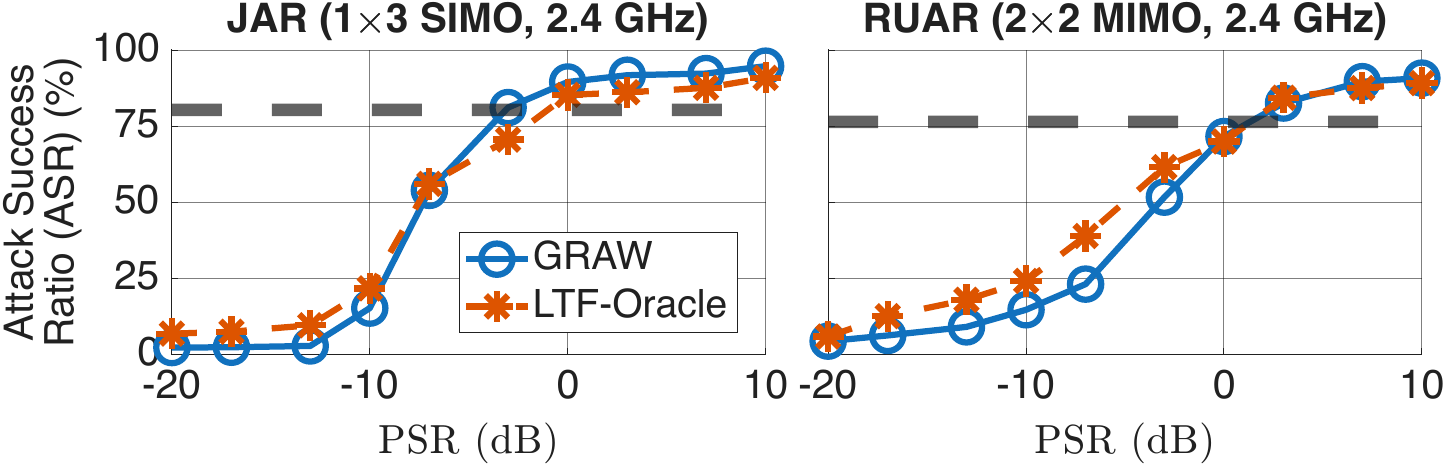}\vspace{-0.3cm}
\caption{Validation of \systemname{}'s multi-antenna LTF manipulation scheme under SIMO and MIMO datasets.}
\label{fig:mimo_ltf}
\end{figure}

\parahead{HAR classifiers.} We assess \systemname{} against architectures matched to the surrogate (Bi-LSTM) and unmatched (CNN, attention, random forest, and transformer) in Table~\ref{tab:HARclass}, assessing cross-architecture transferability. We further test classifiers taking either raw CSI or statistical features as input, including variance and SNR~\cite{jannat2023efficient} or short-time Fourier transform (STFT)~\cite{zhang2023imgfi}.

For \TAR~and \JAR, we train the surrogate classifier and the online perturbation generator on the downsampled CSI. The downsampling lets a fixed input length $\ell$ cover a longer time duration. We heuristically set $\ell=5$ since larger values increase model complexity and degrade training stability, while smaller ones shorten the covered duration and reduce attack performance. Human movement induces Doppler shifts of at most 30~Hz~\cite{kim2009human}, with most activity-relevant energy concentrated at lower frequencies. Sampling in the tens-of-Hz range therefore retains the components needed for activity classification, which we confirm empirically in \S\ref{subsec:validation}. Accordingly, we use 50~Hz for \TAR{} and \RUAR{}, and 40~Hz for \JAR{} (an integer factor of its original 320~Hz rate). Hyperparameters for training the surrogate classifier (Table~\ref{tab:LSTMparam}) and GAIL policy network (Table~\ref{tab:GAILparam}) are also summarized.

\vspace{-7pt}
\subsection{Comparison Schemes}
\label{s:comparision_scheme}
We test three digital-attack schemes (classifier input manipulation) and two remote-attack schemes (LTF manipulation) as comparison schemes:

\parahead{WiCAM.} \WiCAM~\cite{xu2022wicam} is a digital attack that uses an attention-based surrogate model~\cite{li2021two} to select critical subcarriers and time steps, and applies FGM only on those positions. The surrogate expects fixed-length inputs, so CSI sequences for variable durations are resampled to fixed length. We set $t = 0.4$, following the original paper's recommendation.




\parahead{Black-box FGM.} Black-box FGM perturbation~$\hat{\AAA}$ in Eq.~\eqref{eq:blackFGM} is computed using the surrogate model~$\fCsur$. Knowledge of full-activity CSI sequence is assumed.

\parahead{Universal FGM.} Universal FGM is a single fixed perturbation per activity class by averaging black-box FGM perturbations~$\hat{\AAA}$ computed on the adversary's training data~\cite{moosavi2017universal}. All~$\hat{\AAA}$ are resampled to the longest activity duration before averaging, and the average perturbation is then resampled to the action length and used to generate the LTF perturbation. This attack does not require the full-activity CSI sequence, but does assume prior knowledge of the activity duration. For these black-box/universal FGM methods, we also test them for remote attacks but under unrealistic assumptions.

\parahead{C\&W}. \CW{} refers to the remote attack method based on the Carlini \& Wagner scheme proposed in~\cite{li2024practical}. Since \CW{} only takes fixed-length sequences as input and does not consider the MIMO constraint (Table~\ref{tab:relworks}), we manually resample the CSI sequences and adapt them for the MIMO (or SIMO) environments using \systemname{}'s scheme (\S\ref{sec:advMIMO}).

\parahead{Behavioral cloning.} Behavioral cloning~\cite{torabi2018behavioral} is an IL-based remote-attack approach. Like \systemname{}, it uses black-box FGM computed on the recent CSI and the pairs, $(\HHH_i^{\ell}, \hat{\AA}_i)$. A Bi-LSTM classifier $f_{\textrm{BC}}(\HHH_i^{\ell})$ is trained in a supervised learning manner (cf. GAIL in \systemname{}) using the pairs as inputs and labels and is deployed as an online perturbation generator. As in \systemname{}, behavioral cloning operates without knowledge of the target classifier, the new HAR CSI, or activity duration.

\subsection{Experimental \systemname{} Design Validations}
\label{subsec:validation}

\begin{figure}
\centering
    \hfill \begin{subfigure}[b]{1.0\columnwidth}
            \hfill \includegraphics[width=1\linewidth]{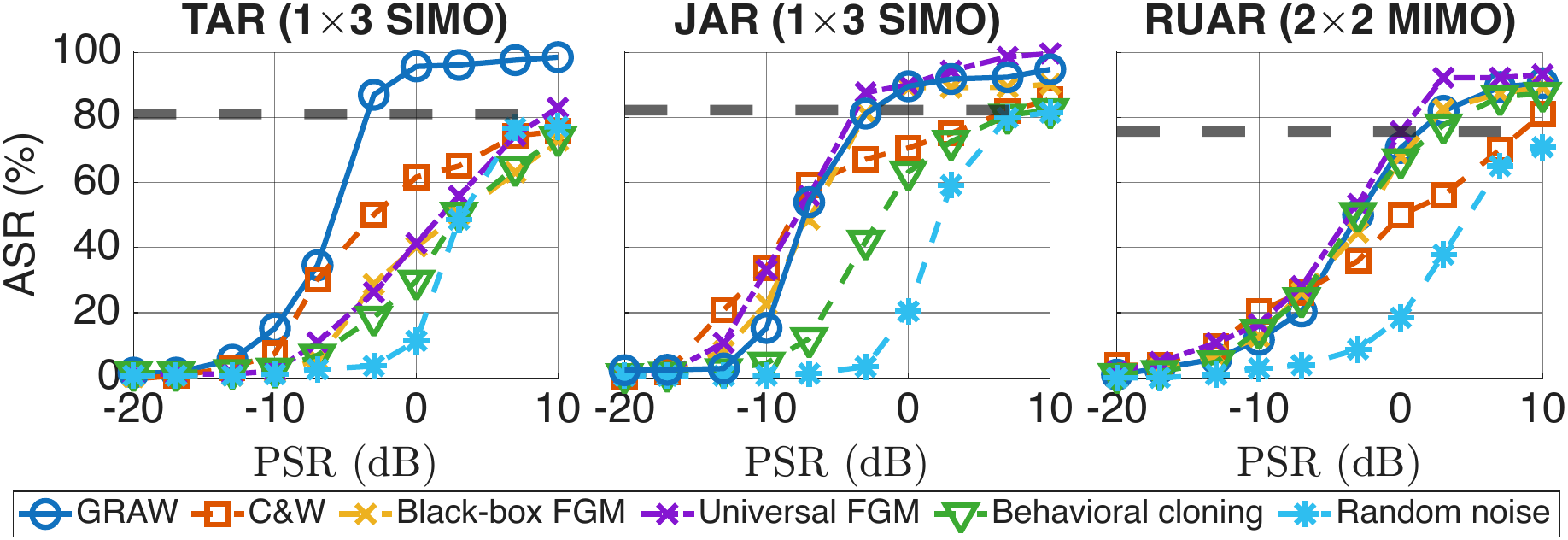}
            \caption{ASR comparison with remote attacks.}
            \label{subfig:ASR_remote}
        \end{subfigure}
    \hfill \begin{subfigure}[b]{\columnwidth}
            \hfill \includegraphics[width=1\linewidth]{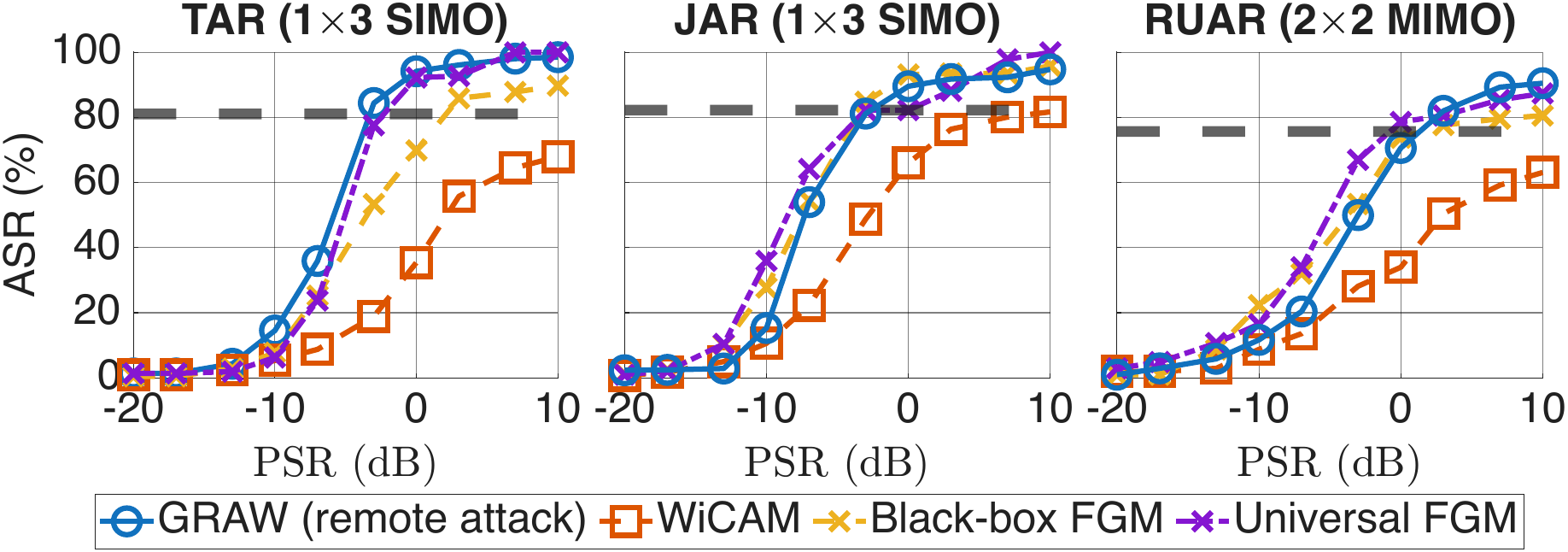}
            \caption{ASR comparison with digital attacks.}\vspace{-0.2cm}
            \label{subfig:ASR_digital}
        \end{subfigure} 
\caption{ASR (Attack Success Ratio) of HAR attacker schemes on the Bi-LSTM-based target classifier across PSRs under various datasets. Dotted lines highlight ASR that makes HAR accuracy random, \emph{i.e.,}  accuracy 1/6 (\TAR{} and \JAR{}) and 1/5 (\RUAR{}). Recall that, except for \systemname{} and behavioral cloning, all the attacker systems have impractical assumptions (\S\ref{s:comparision_scheme}).}
\label{fig:ASR}
\end{figure}

\begin{figure*}
\centering
    \begin{subfigure}[b]{0.99\textwidth}
            \centering 
            \includegraphics[width=0.99\linewidth]{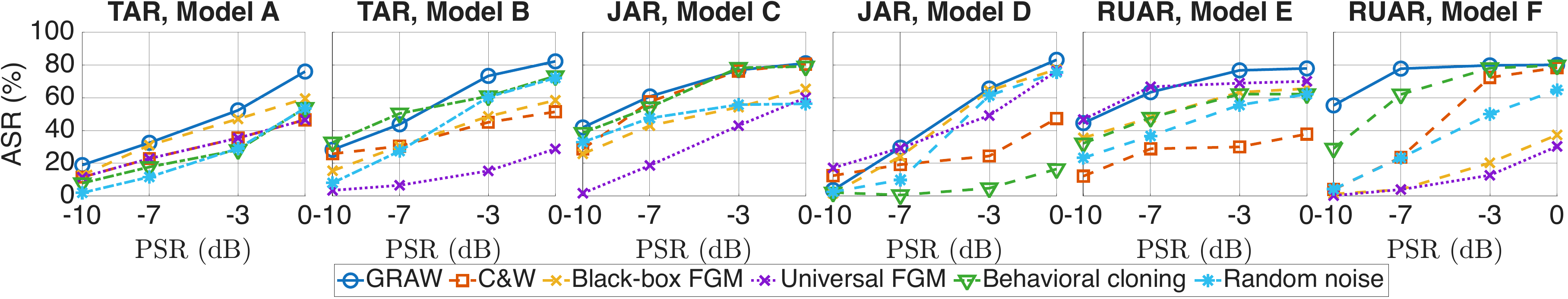}
            \caption{ASR comparison of remote attack schemes with unmatched classifier models (in Table~\ref{tab:HARclass}).}
            \label{subfig:acc_models}
        \end{subfigure}
    
    \begin{subfigure}[b]{0.99\textwidth}
            \hfill \includegraphics[width=1\linewidth]{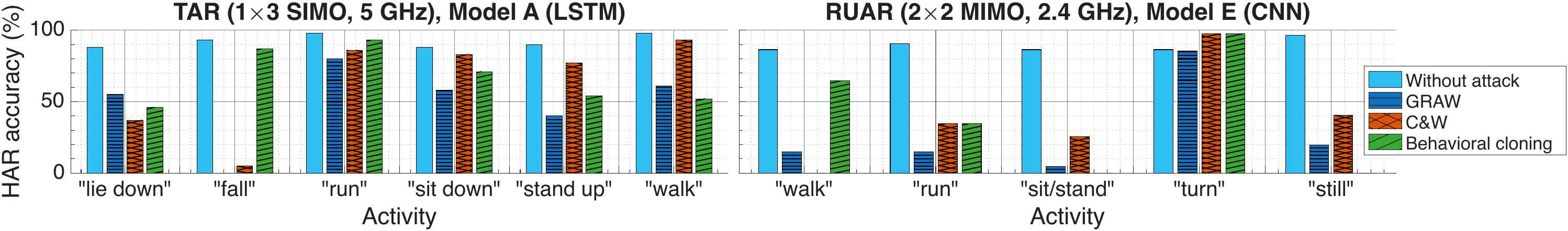}\vspace{-0.2cm}
            \caption{HAR classification accuracy per activity with/without remote attacks (with -3\,dB PSR) under \TAR{} and \RUAR{} datasets.}\vspace{-0.2cm}
            \label{subfig:acc_act}
        \end{subfigure}
\caption{\systemname{} and comparative remote attack scheme results on various target classifiers, where surrogate classifier and HAR classifier models are unmatched (cf. both Bi-LSTM-based classifiers in Figure~\ref{fig:ASR}).}
\label{fig:ASR_models}
\end{figure*}

We validate design components of \systemname{}. We first introduce the evaluation metrics. We use \emph{attack success ratio} (ASR) as the main metric, along with the HAR accuracy. We define $\text{ASR} = (A_0-A)/A_0$, where  $A$ and $A_0$ denote the classifier accuracy with and without attack, respectively. We further introduce \emph{perturbation-to-signal ratio} (PSR), defined as the average amplitude ratio of perturbation  $|\textit{b}_{ij}^k|$ to LTF $|x_{ij}^k|$ across all time steps, subcarriers, and TX antennas, to compare each attack method's degrading efficiency. A desired attack scheme achieves higher ASR at lower PSR.
%

\parahead{Surrogate model.} As the surrogate model is trained on downsampled CSI (\S\ref{subsec:data_target}), we first determine the  downsampling rate by evaluating surrogate accuracy under different rates in Figure~\ref{subfig:surro_dSamp}. The surrogate accuracy deteriorates when the sampling rate is below 10~Hz. This confirms that our chosen rates, 50~Hz for \TAR{} and \RUAR{} and 40~Hz for \JAR, are sufficient to maintain high accuracy. High surrogate accuracy is essential, as it indicates that the surrogate captures HAR-relevant CSI features. The resulting perturbations transfer across diverse target architectures (\S\ref{subsec:att_results}). In our experiments, these models have 93\% average surrogate accuracy across all three datasets (all surrogate accuracies $>$88\% per activity), demonstrating that the Bi-LSTM-based structure can serve as a surrogate for generating reference perturbations.

\parahead{Multi-antenna LTF manipulation.} We validate \systemname{}'s multi-antenna LTF manipulation, by comparing against an \emph{LTF-oracle}, which assumes that each TX-RX antenna pair can use a separate LTF. Figure~\ref{fig:mimo_ltf} plots their ASR across PSR with \TAR{} (SIMO) and \RUAR{} (MIMO) datasets. Gray lines in the figures represent ASR values needed to reduce the target system accuracy to random-selection level. \systemname{} achieves comparable ASR to that of the LTF-oracle across PSR in both SIMO and MIMO settings, with less than 1.9~dB and 0.1~dB PSR difference to achieve 50\% ASR and to degrade the target classifier to random-selection accuracy, respectively, validating the multi-antenna manipulation design.

\parahead{Online amplitude adjustment.} We test whether \systemname{}'s online amplitude adjustment mechanism (\S\ref{sec:ampAdjust}) ensures that the perturbation signals closely match their target amplitude ratio throughout the activity duration. The left panel of Figure~\ref{subfig:ampAdjust} shows the vector norm of adjusted perturbation signals $\BB_i$ of Eq.~\eqref{eq:ampAdjust} (\systemname{}), and the scaled output of Eq.~\eqref{eq:advMIMO} $\gamma\cdot \bar{\BB}_i$, which is the desired result (Oracle). The middle and right panels illustrate $\BB_i$ and $\gamma\cdot \bar{\BB}_i$ at first and last time step, respectively. Although they differ at the first time step due to insufficient history for the amplitude estimator, the two converge over time and become nearly identical at the last time step. This convergence is enabled by the flat temporal amplitude distribution of $\bar{\BB}_i$, which makes $\|\bar{\BBB}_{1:i}\| / \|\textbf{x}_{1:i}\|$ a reliable estimator of the amplitude ratio over the entire activity duration. Across all datasets, \systemname{}'s online adjustment yields an average mismatch between \(\BBB_{1:M}\) and \(\gamma\,\bar{\BBB}_{1:M}\) of less than 5\% over the activity duration (\TAR: 3.5\%, \JAR: 4.9\%, \RUAR: 3.8\%). This indicates that the amplitude adjustment effectively approximates the perturbation scaled by the target amplitude using only recent CSI.

\vspace{-7pt}
\subsection{Adversarial Attack Performance}
\label{subsec:att_results}

Figure~\ref{subfig:ASR_remote} shows the ASR under various remote attack schemes, including \systemname{}. The target model employs an identical Bi-LSTM architecture and sampling rate as the surrogate, representing a matched model scenario. Gray lines indicate the ASR at which the target classifier is degraded to random-selection accuracy, as in Figure~\ref{fig:mimo_ltf}. Across all datasets, \systemname{} achieves performance comparable to or better than the comparison methods, despite its zero-knowledge operation. Notably, in \TAR{}, \systemname{} reaches 50\% and 80\% ASR with 2\,dB and 11\,dB lower PSR, respectively, than the second-best scheme.

The ASR compared to digital attack schemes is plotted in Figure~\ref{subfig:ASR_digital}. Across all the datasets, \systemname{} matches the accuracy degradation of digital attacks with a comparable PSR, requiring 1.6\,dB and 1.9\,dB higher PSR for 50\% and 80\% ASR in the worst cases, respectively. These results are notable, given that \systemname{} operates entirely OTA, without the new HAR CSI access that digital attacks require.

\parahead{Cross-Model Evaluation.} We also test \systemname{}'s transferability to target architectures that differ from its Bi-LSTM surrogate. In Figure~\ref{subfig:acc_models}, \systemname{} consistently delivers the highest or near-highest ASR across all PSR values. While \CW{} occasionally approaches \systemname{} (\emph{e.g.,} Model~C), but is unstable elsewhere. Universal FGM shows similar inconsistency, occasionally matching \systemname{} (e.g., Model~E) but often collapsing (e.g., Model~F), as its fixed per-class perturbation cannot adapt to the target. \systemname{} remains effective against attention-based architectures---the attention model (Model~B) and the Transformer (Model~F)---achieving the highest ASR especially at low PSR, despite using only a Bi-LSTM surrogate. Overall, the results highlight the robustness of \systemname{} across diverse HAR models, maintaining strong performance.

We further analyze per-activity accuracy for \systemname{}, behavioral cloning, and \CW{} in Figure~\ref{subfig:acc_act}. Interestingly, the impact of perturbation varies substantially across activities. For example, against Model~A, perturbations with a PSR of -3\,dB drive the accuracy on ``fall'' data nearly to 0\%, while the accuracy on ``run'' data drops far less. In \RUAR{}, most perturbed CSI sequences are misclassified as ``turn'', revealing that perturbations tend to drive the classifier toward a dominant decision region. Even so, the average accuracy across activities falls to the random-selection level. This activity-dependent vulnerability suggests that \systemname{} could be further strengthened by adaptively tuning the PSR per activity.

\begin{figure}
\centering
    \hfill \begin{subfigure}[b]{0.50\columnwidth}
            \hfill \includegraphics[width=1\linewidth]{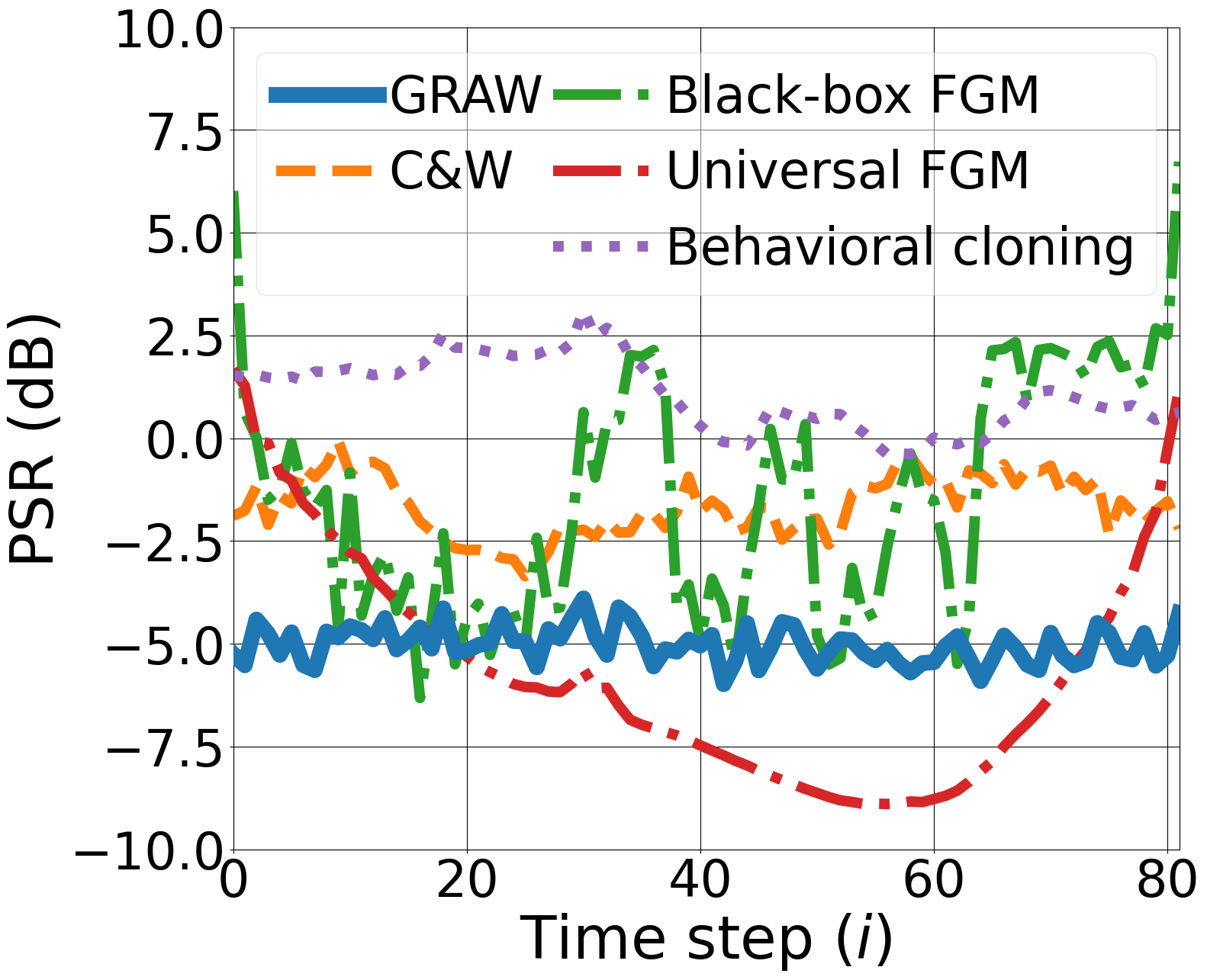}\vspace{-0.1cm}
            \caption{``stand up'' data in \TAR}\vspace*{-0.2cm}
            \label{subfig:ampRatioDist_TAR}
        \end{subfigure}
    \hfill \begin{subfigure}[b]{0.45\columnwidth}
            \hfill \includegraphics[width=1\linewidth]{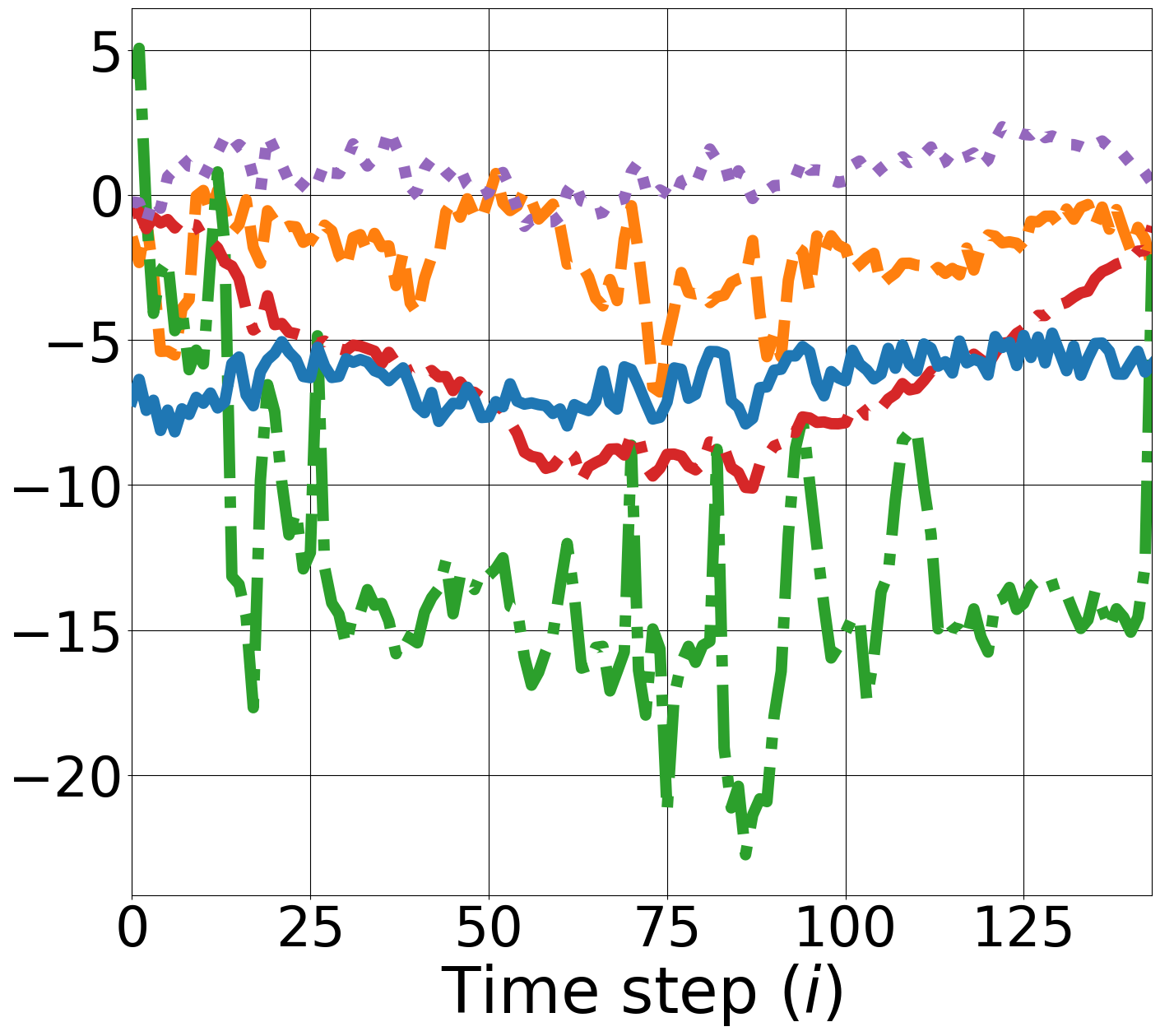}\vspace{-0.1cm}
            \caption{``turn'' data in~\JAR}\vspace*{-0.2cm}
            \label{subfig:ampRatioDist_JAR}
        \end{subfigure}
\caption{Required PSR over time ($r_i$) to reduce Bi-LSTM classifier HAR accuracy to a random-selection level. Compared to other designs, \systemname{} requires relatively consistent and low PSRs (\S\ref{sec:ampAdjust}).}
\label{fig:ampRatioDist}
\end{figure}

The efficiency of \systemname{} across diverse target model structures is attributed to the temporal distribution of perturbation signal power. To analyze this, we compute the perturbation amplitude ratio per time step $r_i$, averaging $|b_{ij}^k|/{|x_{ij}^k|}$ over subcarrier $j$ and TX antenna $k$.
Figure~\ref{fig:ampRatioDist} illustrates $r_i$ over time for one ``stand up'' sample in~\TAR~and ``turn'' sample in~\JAR~under PSR values configured to degrade the target HAR accuracy to random-selection level. \systemname{} and behavioral cloning generate flatter and more consistent amplitude distributions than other methods, as they produce outputs based only on the past $\ell$ steps and thus cannot significantly amplify specific temporal segments. In contrast, the other methods compute their perturbations by processing the CSI over entire activity at once, making them concentrate perturbation power on critical time steps (e.g., universal FGM at $40 \le i \le 60$ in Fig.~\ref{subfig:ampRatioDist_TAR}). Target classifiers such as~\cite{chen2018wifi, islam2022stc} in Table~\ref{tab:HARclass} may take only time indices with low amplitudes (e.g., black-box FGM at $75 \le i \le 125$ in Fig.~\ref{subfig:ampRatioDist_JAR}) as input, thereby diminishing attack effectiveness. However, the flat power distribution of \systemname{} avoids this limitation, efficiently degrading target HARs regardless of their input timing.

\begin{figure}
\centering
    \includegraphics[width=\linewidth]{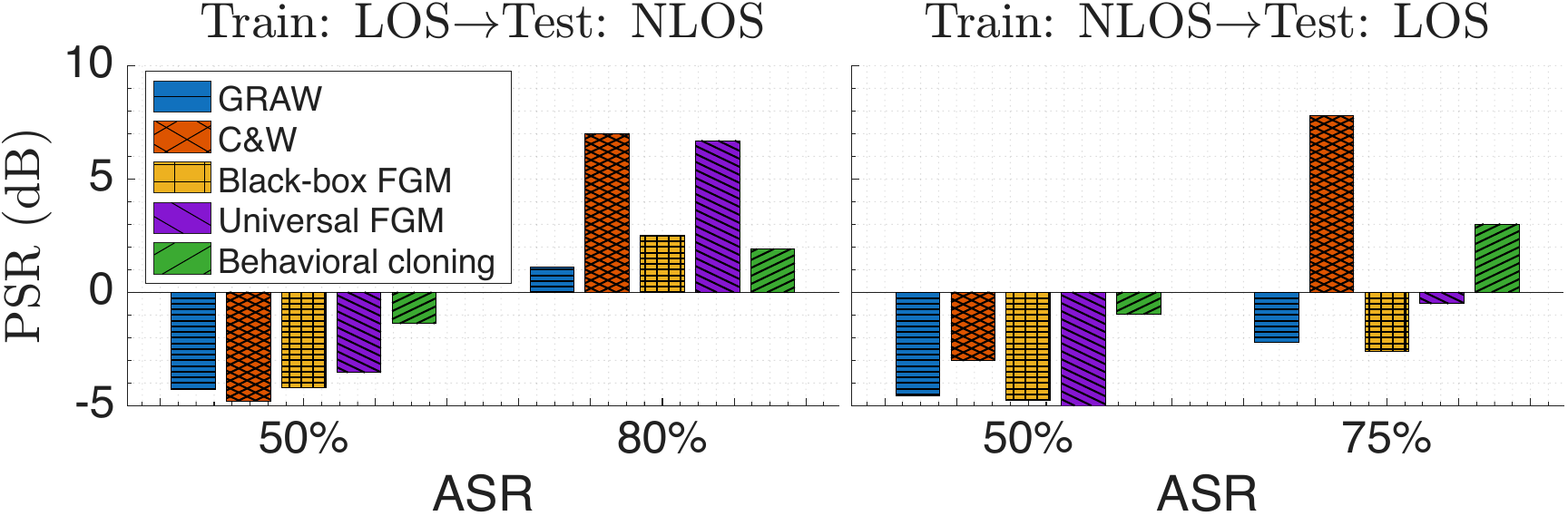}\vspace*{-0.3cm}
\caption{Cross-environment evaluations with required PSR to degrade Bi-LSTM classifier HAR accuracy to ASR (50\%, 80\% (LOS $\rightarrow$ NLOS), 75\% (NLOS $\rightarrow$ LOS)) on the \JAR{} dataset.}
\vspace{-0.3cm}
\label{fig:targetAcc}
\end{figure}

\parahead{Cross-Environment Evaluation.} Figure~\ref{fig:targetAcc} reports the PSR values needed to degrade the target classifier accuracy to 50\% and random-selection level ASR (80\% for line-of-sight (LOS)$\rightarrow$NLOS, 75\% for NLOS$\rightarrow$LOS) when training and test environments differ on the \JAR{} dataset. When the surrogate is trained on LOS and tested on NLOS (Fig.~\ref{subfig:env_JAR}), \systemname{} achieves both ASR targets with PSR comparable to black-box FGM. By contrast, \CW{} and universal FGM, which apply perturbations computed on training data without adaptation, require over 5\,dB PSR to reach random-selection level ASR. In the reverse NLOS$\rightarrow$LOS setting, \systemname{} reaches the 75\% ASR at -3\,dB PSR, whereas behavioral cloning needs 3\,dB. This robustness stems from GAIL's distribution matching, which generalizes better than behavioral cloning's pointwise state-action fitting (\S\ref{subsec:GAIL}).

\vspace{-6pt}
\subsection{Impact on Data Communication Link}
\label{s:impact_link}


Since communication is the primary goal of Wi-Fi, the transmit perturbation signal aims to degrade the classifier without disrupting communication. This subsection evaluates \systemname{}'s impact on communication performance. Note that measured bit error rate (BER) and packet success rate results appear in \S\ref{subsec:real-time}, as measuring them requires payload transmission (beyond channel datasets in \TAR{} and \JAR{}). 

Even at the same PSR, different perturbation amplitude distributions can lead to different effects on the link. To capture this, we quantify the interference using estimated throughput based on per-subcarrier \emph{signal-to-perturbation-and-noise ratio} (SPNR) over time, $\bar{r}_{ij}$, assuming maximum ratio transmission (MRT) and maximum ratio combining (MRC) for TX and RX sides, respectively, as follows. First, we compute SPNR under a chosen SNR, $r_{ij}^k = {|x_{ij}^k|^2}/({|b_{ij}^k|}^2+|n_{ij}^k|^2)$, where $n_{ij}^k$ depends on the SNR. Next, we sum $r_{ij}^k$ across TX antennas $k$ in the linear domain and add $10\log_{10}(\nRX)$ to account for receiver diversity. Then, we map the resulting SPNR, $\bar{r}_{ij}$, to the highest modulation and coding scheme (MCS) supported, according to the Wi-Fi standard table~\cite{ieee802.11-2024}.

\begin{figure}
    \centering
            \includegraphics[width=0.94\linewidth]{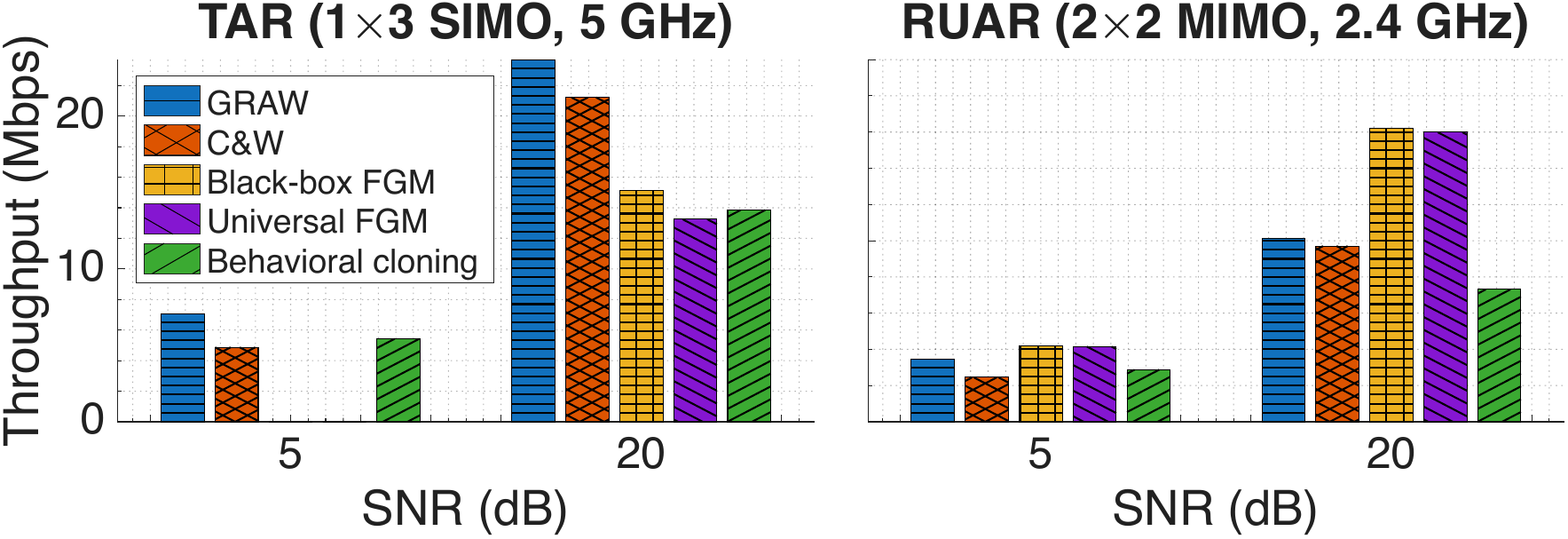}\vspace*{-0.3cm}
            \label{subfig:thrpt_snr}
\caption{Estimated throughput across SNRs of Wi-Fi data communication under attacker systems with \TAR{} and \RUAR{} datasets, where PSR is set to degrade the target classifier to random-selection level. Spatial diversity MIMO is used.}
\label{fig:MIMO_impact}
\end{figure}

Figure~\ref{fig:MIMO_impact} shows the average estimated throughput, when PSR is set to degrade the target classifier to random-selection level. In \TAR{}, \systemname{} achieves the highest estimated throughput, as it requires smaller PSR than other schemes. At 5\,dB SNR, black-box and universal FGM nearly eliminate data communication, with throughput approaching 0\,Mbps. In \RUAR{}, \systemname{} achieves lower throughput than black-box and universal FGM because \systemname{} requires a higher PSR to reach random-selection level in this dataset. Nevertheless, \systemname{} still outperforms \CW{} and behavioral cloning, which do not assume full-activity CSI sequence.

\begin{figure}
\centering
\includegraphics[width=\columnwidth]{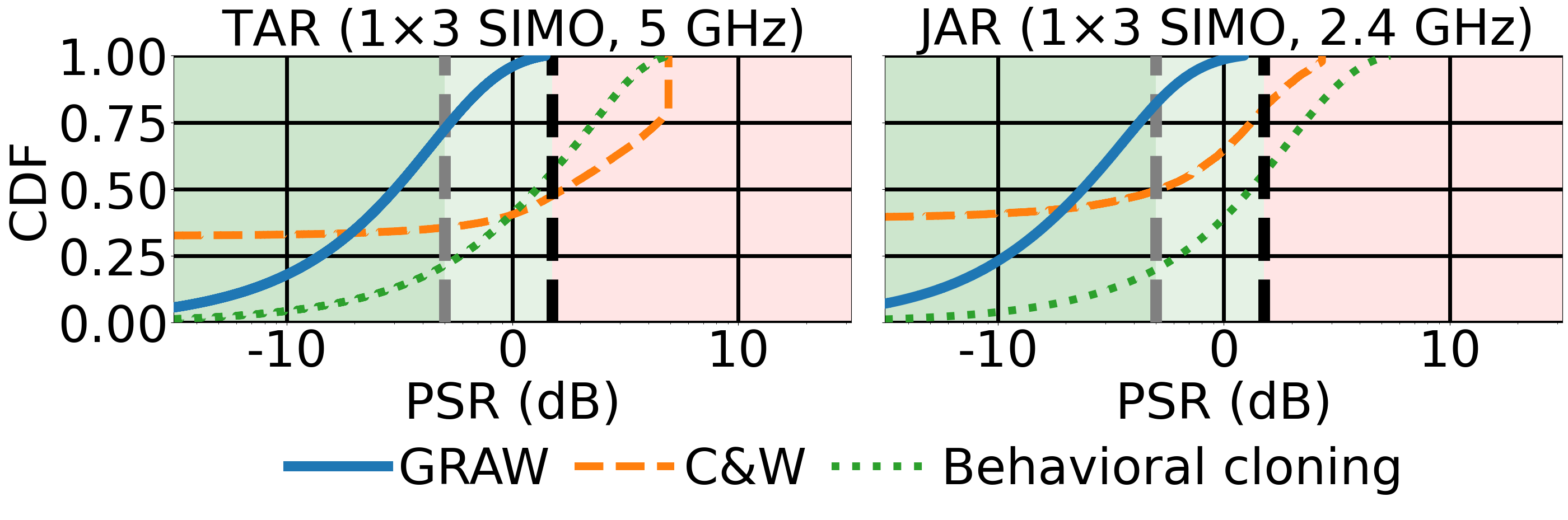}\vspace*{-0.3cm}
\caption{CDF of per-subcarrier PSR values over time required to degrade the target classifier to random-selection level. The vertical dashed lines indicate PSR thresholds for reliable Wi-Fi communication without MRC (gray) and with MRC (black).}
\label{fig:cdf_psr}
\end{figure}

Figure~\ref{fig:cdf_psr} shows the CDF of per-subcarrier PSR values over time. The gray (\emph{left}) and black vertical dashed lines (\emph{right}) indicate PSR thresholds for reliable communication without and with MRC, respectively; green and red regions indicate reliable and unreliable zones. Perturbations whose PSR exceeds these thresholds disrupt link availability due to the channel distortion. With MRC, \systemname{} keeps all per-subcarrier PSR values below the threshold (0.0\%) in both datasets, while \CW{} and behavioral cloning exceed it by 52.5\%/19.7\% and 43.6\%/44.2\%, respectively; without MRC, \systemname{} still maintains a lower fraction than competing schemes. This stems from two factors: \systemname{}'s smaller PSR (Fig.~\ref{subfig:ASR_remote}) and its flatter amplitude distribution (Fig.~\ref{fig:ampRatioDist}), which avoids spikes that push per-subcarrier PSR above the threshold.

\vspace{-0.3cm}
\subsection{Real-Time Over-the-Air Experiments}
\label{subsec:real-time}

We also evaluate \systemname{} on SDRs in a real-time, OTA manner to demonstrate its deployment feasibility. We implement the inference stage as an out-of-tree GNU Radio~\cite{gnuradio} block in C++, given that training is performed offline. The block performs channel estimation from received LTF signals, computes the perturbation using the trained policy network, and transmits the modified LTF in real-time on our SDR platforms (\RUAR{}). The inference latency of 0.6~ms on CPUs is well below the 20~ms sampling period, confirming hardware timing feasibility without any modifications.

\begin{figure}
\centering
\begin{subfigure}[b]{.95\columnwidth}
            \centering \includegraphics[width=\linewidth]{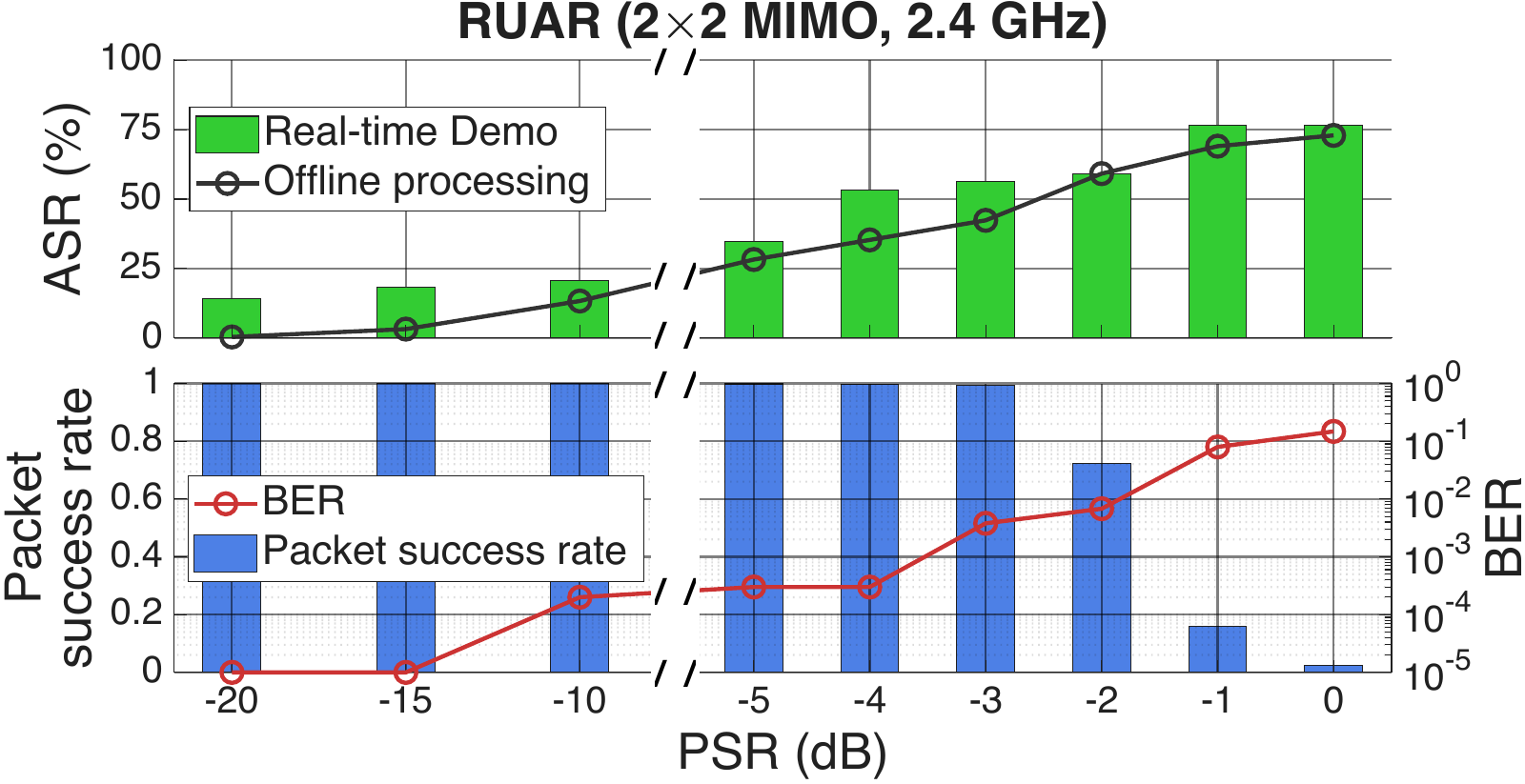}\vspace{-0.2cm}
            \caption{ASR, BER, and packet success rate across PSRs.}
            \label{subfig:demo_ASR_BER}
        \end{subfigure}
        \\
        \begin{subfigure}[b]{0.95\columnwidth}
            \centering \includegraphics[width=\linewidth]{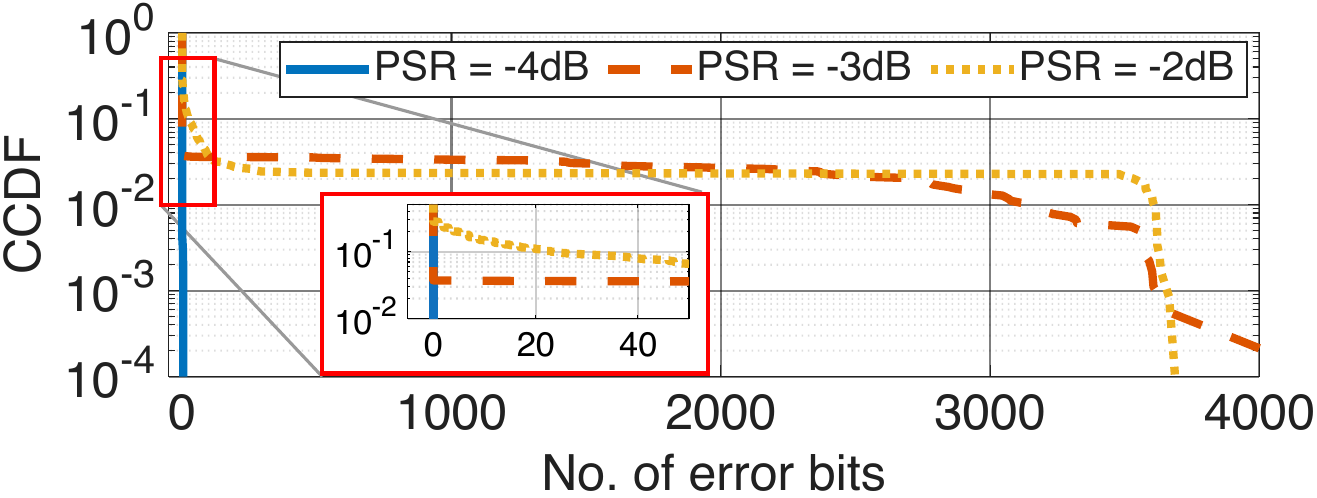}\vspace{-0.2cm}
            \caption{Per-packet BER distribution. Most of the errors are observed in specific packets (\emph{i.e.,} burst error).}\vspace{-0.2cm}
            \label{subfig:hist_BER}
        \end{subfigure}
\caption{OTA, real-time demonstration of \systemname{} under spatial multiplexing MIMO, measured under \RUAR{} setting in Figure~\ref{fig:env}.}
\label{fig:demo}
\vspace{-0.30cm}
\end{figure}

\parahead{Attack performance.} The upper panel of Figure~\ref{subfig:demo_ASR_BER} shows that the real-time demo follows a similar trend to offline processing across PSR values. In particular, at low PSR, the real-time demo achieves higher ASR than offline processing. Minor differences arise from analog and PHY effects in the RF pipeline (e.g., DAC non-linearities, CFO) absent in offline processing. Understanding how RF chains affect perturbation may provide insights for future perturbation designs.

\parahead{Impact on data communication link: BER and packet success rate.} The lower panel of Figure~\ref{subfig:demo_ASR_BER} shows BER and packet success rate measured in a spatial multiplexing MIMO environment. At PSR = -4\,dB and -3\,dB, \systemname{} achieves over 50\% ASR while maintaining a 99.7\% and 96.3\% packet success rate, respectively, confirming its minimal degradation on regular Wi-Fi data communications. With the diversity gain from MRT and MRC, this reliability extends to higher perturbation levels (PSR = 2\,dB), beyond the point where \systemname{} already reaches the random-selection level (PSR = -2\,dB). Interestingly, at PSR = -4\,dB and -3\,dB, the measured BER of $3\times10^{-4}$ and $6\times10^{-4}$ would suggest a much lower packet success rate than the observed 99.7\% and 96.3\% if the errors were uniformly distributed over 1000-byte (8000-bit) packets. We found that this discrepancy arises because perturbation signals corrupt channel estimation, causing error bits to concentrate in specific packets, as shown in Figure~\ref{subfig:hist_BER}. While most packets are zero-error, the remaining packets suffer severe channel estimation errors, and thus most of their following data payload bits are in error. This is directly observed in the figure with rapid drops in the CCDF around 3700--4000 bit errors out of 8000 bits (Fig.~\ref{subfig:hist_BER}).


\parahead{Power Consumption.} We evaluate \systemname{}'s power consumption (\S\ref{subsec:GAIL}, \S\ref{sec:advMIMO}, and \S\ref{sec:ampAdjust}) via FLOP analysis. Generating a perturbation signal for one LTF requires \(2.2\times10^5\) FLOPs, or \(1.1\times10^8\) FLOPs/s at the 50~Hz update rate. Assuming an energy of \(\approx\) 0.6\,pJ/FLOP for an RTX-4080Ti GPU~\cite{RTX-4080Ti} and \(\approx\) 0.9\,nJ/FLOP for a Cortex-A53 CPU~\cite{archer-ax1800}, commonly used in commodity Wi-Fi routers, the additional power cost is about 0.067\,mW on the GPU and 0.1\,W on the CPU, below 1\% of the maximum power consumption of the GPU and a commodity Wi-Fi router, 450~W and 12~W.

\begin{table}[t]
\begin{tiny}
\centering
\caption{Estimated power consumptions}\vspace*{-0.2cm}
\label{tab:energy}
\begin{tabular}{c|cccccc}
\toprule
\makecell{\textbf{Power}\\\textbf{consumption}} & 
\makecell{\systemname{}} & 
\makecell{\textbf{Behavioral}\\\textbf{cloning}} & 
\makecell{\CW} & 
\makecell{\textbf{Black-box}\\\textbf{FGM}} & 
\makecell{\textbf{Universal}\\\textbf{FGM}} & 
\makecell{\WiCAM} \\ \midrule
\makecell{\textbf{(mW)}}  & 11 & 10 & 0.73 & 560 & 30 & 1140  \\ 
\bottomrule
\end{tabular}
\vspace{-0.45cm}
\end{tiny}
\end{table}
\setlength{\tabcolsep}{\oldtabcolsep} 

Table~\ref{tab:energy} summarizes the estimated power consumption of \systemname{} and comparison schemes. \systemname{} consumes only 11\,mW, comparable to behavioral cloning. Gradient-based methods that compute perturbations at runtime (Black-box FGM and \WiCAM) require far more power (560 and 1140\,mW), while Universal FGM (30\,mW), which reuses precomputed per-class perturbations, and \CW{} (0.73\,mW) remain low. 

\parahead{Memory footprint.} The perturbation generator is a four-layer MLP with 0.19--0.33\,M parameters, occupying 0.7--1.3\,MB in FP32, a small fraction of the RAM on commodity APs (typically tens to hundreds of MB). Thus \systemname{} fits comfortably alongside routing and NAT tables.

\vspace{-5pt}
\section{Discussion}
\label{sec:disc}

\parahead{Practical deployments.} 
PHY-level operations such as LTF manipulation are inaccessible on commodity Wi-Fi chipsets; even open-source Wi-Fi router firmware (\emph{e.g.}, OpenWRT~\cite{OpenWRT}, DD-WRT~\cite{DD_WRT}) does not expose them. We therefore envision chipset and router vendors natively integrating \systemname{} into their products as a privacy-protection feature. Although this vendor-level integration may appear to limit immediate deployability, this constraint is appropriate for our privacy objective. The PHY-level security functionality should remain within the trusted router hardware; exposing it to untrusted user-space software or external devices could introduce another privacy and security risk. Unlike denial-of-service jamming~\cite{liu2023time}, \systemname{} modifies only the router's own transmission without blocking other devices' channel access.

\parahead{Opportunistic operation.} 
\systemname{} has reduced the negative impact on the communication link, but its perturbations noticeably degrade link quality for high ASRs.
Therefore, the \systemname{} active duration should be minimized; it should operate only opportunistically rather than continuously, for example, by activating perturbations only when a malicious user device running a HAR application is detected. 
However, such detection is currently infeasible, as HAR runs entirely within the user device. Still, one might \emph{approximate} the unauthorized HAR activity state, \emph{e.g.,} monitoring indirect signals such as periodic CSI-request patterns. Given the growing importance of privacy in Wi-Fi HAR, automated detection of unauthorized HAR will become increasingly essential. Furthermore, the co-existence of legitimate and malicious devices would present an interesting scenario. In such mixed environments, the router must target only malicious HAR while avoiding unnecessary degradation for legitimate HAR devices. We will explore these challenges in our future work.

\begin{figure}
\centering
    \includegraphics[width=.95\linewidth]{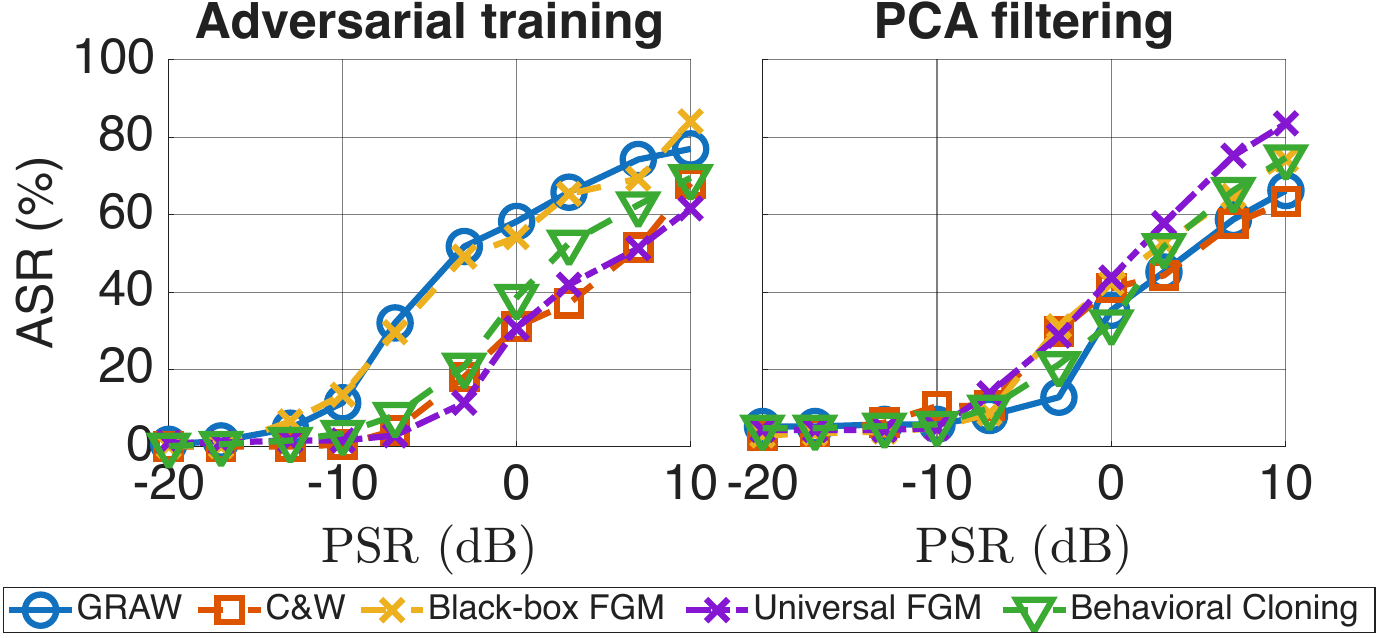}\vspace{-0.3cm}
\caption{ASR across PSR against a Bi-LSTM-based HAR classifier with defense systems (adversarial training and PCA filtering) on the \TAR{} dataset.}
\vspace{-0.3cm}
\label{fig:ASR_defense}
\end{figure}

\parahead{Robustness against defense algorithms.}
Recently, defense systems against adversarial attacks have also emerged, generally aiming to either mitigate or withstand (malicious) perturbations. In this context, we also test \systemname{} against two representative defense mechanisms, adversarial training~\cite{kurakin2016adversarial} and principal component analysis (PCA) based defense~\cite{bhagoji2018enhancing}. As shown in Figure~\ref{fig:ASR_defense}, under adversarial training, \systemname{} outperforms behavioral cloning and \CW{}. Under the PCA filtering, however, behavioral cloning achieves the highest ASR, while \systemname{} experiences a larger drop in ASR. We will explore advanced \systemname{} designs in our future work, targeting malicious HAR devices with such defense systems. 

\parahead{Other possible applications.}
Beyond Wi-Fi HAR, remote adversarial attacks have been explored in various neural network–based wireless applications.
Perturbation signals are designed to attack DL-based modulation classification~\cite{wang2024wireless}, autoencoder-based decoders~\cite{chang2023magmaw},
device-identification systems, and DL-based indoor localization~\cite{liu2023exploring}. However, these methods~\cite{xiao2023over, wang2024wireless} often assume perfect synchronization between the adversary's and the target system's inputs, making the practical deployment infeasible. 
Therefore, the \systemname{}-based approach can be adapted and applied to these applications, to enable the adversarial attacks without requiring synchronization (or any information on the target system). 

\vspace{-5pt}
\section{Conclusion}
\label{sec:conc}
This paper presents \systemname, a remote adversarial attacker system against Wi-Fi-based HAR. To our best knowledge, this is the first application of GAIL to remote adversarial attacks against Wi-Fi HAR and thus completely eliminates the need for information on the target HAR systems.
\systemname{} also resolves a constraint in multi-antenna environments and maintains a consistently low perturbation signal power, thus minimizing its negative impact on regular Wi-Fi data communication at the same time. 
Given the impressive performances observed, we believe this GAIL-based zero-knowledge adversarial attack approach is worth exploring further, not just for Wi-Fi HAR, but also for a variety of applications such as 
mmWave radar and wireless indoor localization. 

\clearpage

\bibliographystyle{ACM-Reference-Format}
\bibliography{bibdata}

\appendix

\section{TRPO detailed steps}

\label{app:TRPO}
Policy gradient aims to increase the probabilities of actions that yield higher returns along learner trajectories. The GAIL training process includes a policy gradient to align the policy function closely with expert trajectories. We implement trust region policy optimization (TRPO) as our policy gradient algorithm, which optimizes two key components; the policy function $\pi(\textbf{A}_i|\HHH_i^{\ell})$ and the value function, $V(\HHH_i^{\ell})$. Following~\cite{schulman2015trust}. For value function estimation, we employ generalized advantage estimation (GAE)~\cite{schulman2015high}. In $k$th iteration of GAIL, the detailed TRPO steps executed in line 4 in Algorithm~\ref{alg:GAILattack} are as follows:
\begin{enumerate}
\item{Compute temporal difference (TD) error: 

$\delta_i^{V_{\phi_k}}=-C(\HHH_i^{\ell},\textbf{A}_i) + \gamma V_{\phi_k}(\HHH_{i+1})-V_{\phi_k}(\HHH_i^{\ell})$ at all time steps $i\in\{1,2,\cdots,M\}$}
\item{
Compute advantage values: $\hat{A}_i=\sum_{l=0}^{\infty}{(\gamma \lambda_G)^l\delta_{m+l}^{V_{\phi_k}}}$ at all time steps $i\in\{1,2,\cdots,M\}$}
\item{
Update the parameters $\phi$ of value function $V_{\phi}(\HHH_i^{\ell})$ to decrease the objective, $K(\phi)$: $\phi_{k+1}\leftarrow \phi_k-\nabla _{\phi} K(\phi)$
$$
\begin{aligned}
K(\phi)=&\sum_{i=1}^{M} ||V_{\phi}(\HHH_i^{\ell})-\hat{V}_i||^2 \\
\textrm{subject to }& \frac{1}{M}\sum_{i=1}^{M}\frac{||V_{\phi}(\HHH_i^{\ell})-\hat{V}_i||^2}{2\sigma^2} \le \epsilon_K, \\
\hspace{20pt}\textrm{where }\hat{V}_i=\sum_{l=0}^{\infty}\gamma^lr_{i+l}&\textrm{ and }\sigma^2=\frac{1}{M}\sum_{i=1}^{M}{||V_{\phi_k}(\HHH_i^{\ell})-\hat{V}_i||^2}.
\end{aligned}
$$}

\item{Update the parameters $\theta$ of policy function $\pi_{\theta}$ to decrease the objective, $L_{\theta_k}(\theta)$: $\theta_{k+1}\leftarrow \theta_k - \nabla_\theta L_{\theta_k}(\theta)$
$$\begin{aligned}
\hspace{-30pt}\textrm{subject to}& \textrm{ KL divergence, } \overline{D}_{\textrm{KL}}^{\theta_{k}} (\pi_{\theta_k}, \pi_\theta )\le \epsilon_{\pi}\\
    \textrm{where }L_{\theta_k}(\theta)&=\frac{1}{M}\sum _{i=1}^M {\frac{\pi_\theta(\textbf{A}_i|\HHH_i^{\ell})}{\pi_{\theta_k}(\textbf{A}_i|\HHH_i^{\ell})}} \hat{A}_i\\
    \overline{D}_{\textrm{KL}}^{\theta_k} (\pi_{\theta_k}, \pi_\theta)& = \frac{1}{M}\sum _{i=1}^ND_{\textrm{KL}}(\pi_{\theta_k}(\cdot | \HHH_i^{\ell}) || \pi_{\theta}(\cdot | \HHH_i^{\ell})).
\end{aligned}$$}
\end{enumerate}

\section{Neural network parameters}

\begin{table}[H]
\centering
\begin{footnotesize}
\caption{Surrogate Bi-LSTM HAR classifier model parameters}
\label{tab:LSTMparam}
\begin{tabular}{c|c|c|c}
\toprule
\textbf{Parameter} & \textbf{Value} & \textbf{Parameter} & \textbf{Value} \\ \midrule
\multicolumn{1}{c|}{\begin{tabular}[c]{@{}c@{}}\textbf{LSTM hidden}\\\textbf{layer dimension}\end{tabular}}& 200 &\multicolumn{1}{c|}{\begin{tabular}[c]{@{}c@{}}\textbf{Learning}\\\textbf{rate}\end{tabular}} & $1\times10^{-5}$\\ \hline
\textbf{Loss function} & \multicolumn{1}{c|}{\begin{tabular}[c]{@{}c@{}}Cross-\\entropy\end{tabular}} & \textbf{Epoch} & 400 \\ 
\bottomrule
\end{tabular}
\end{footnotesize}
\end{table}

\begin{table}[H]
\caption{GAIL network parameters}
\label{tab:GAILparam}
\centering
\begin{footnotesize}
\begin{tabular}{c|c||c|c}
\toprule
\textbf{Parameter} & \textbf{Value} & \textbf{Parameter} & \textbf{Value}\\ \midrule
\textbf{Epochs} & 4000 & \textbf{$D_w$ learning rate} & $2\times10^{-5}$ \\ \hline
\multicolumn{1}{c|}{\begin{tabular}[c]{@{}c@{}}\textbf{Input}\\\textbf{length ($\ell$)}\end{tabular}}& 5 & {\begin{tabular}[c]{@{}c@{}}\textbf{Hidden layer}\\\textbf{dimension}\end{tabular}} & 200 \\ \hline
$\epsilon_K$, $\epsilon_{\pi}$ & 0.01 & \textbf{No. hidden layers} & 4 \\ \hline
\multicolumn{1}{c|}{\begin{tabular}[c]{@{}c@{}}\textbf{Discount}\\\textbf{factor} ($\gamma$)\end{tabular}} & 0.99 &\multicolumn{1}{c|}{\begin{tabular}[c]{@{}c@{}}\textbf{Policy regularizer}\\\textbf{coefficient} ($\lambda_G$)\end{tabular}} & 0.01\\
\bottomrule
\end{tabular}
\end{footnotesize}
\end{table}

\end{document}